\documentclass[english,11pt]{article}
\usepackage{lmodern}

\usepackage[T1]{fontenc}
\usepackage[latin9]{inputenc}
\usepackage{prettyref}
\usepackage{booktabs}
\usepackage{amsmath}
\usepackage{amssymb}
\usepackage{graphicx}

\makeatletter

\newcommand{\lyxaddress}[1]{
\par {\raggedright #1
\vspace{1.4em}
\noindent\par}
}

\usepackage{mathrsfs}   %\mathscr{L}
\usepackage{slashed}     %\slashed{p}
\usepackage{bbold}  % \mathbb{1} for the identity matrix
\usepackage{url}
\usepackage{graphicx}
\usepackage[colorlinks=true,linkcolor=redLinks,citecolor=greenLinks,urlcolor=redLinks, pdfborder={0 0 1}]{hyperref}
\usepackage{color} 
\usepackage[numbers,sort&compress]{natbib}

\definecolor{greenLinks}{rgb}{0, 0.6, 0} 
\definecolor{blueLinks}{rgb}{0, 0, 0.6}
\definecolor{redLinks}{rgb}{0.6, 0, 0}
\definecolor{eprintLinks}{rgb}{0.4, 0.4, 0.4}
\definecolor{journalLinks}{rgb}{0.6, 0, 0}

\newcommand{\MYhref}[3][redLinks]{\href{#2}{\color{#1}{#3}}}%

\usepackage{multirow}
\let\orig@Hy@EveryPageAnchor\Hy@EveryPageAnchor
\def\Hy@EveryPageAnchor{%
    \begingroup
    \hypersetup{pdfview=Fit}%
    \orig@Hy@EveryPageAnchor
    \endgroup
}

\let\oldFootnote\footnote
\newcommand\nextToken\relax

\renewcommand\footnote[1]{%
    \oldFootnote{#1}\futurelet\nextToken\isFootnote}

\newcommand\isFootnote{%
    \ifx\footnote\nextToken\textsuperscript{,}\fi}

\global\long\def\rep#1{\boldsymbol{#1}}
\global\long\def\repb#1{\left(\boldsymbol{#1}\right)}

\makeatother

\usepackage{babel}
\begin{document}

\title{{\Large{}\vspace{-1.0cm}} \hfill {\normalsize{}IFIC/15-24} \\*[10mm] On
the chirality of the SM and the fermion content of GUTs {\Large{}\vspace{0.5cm}}}

\author{{\Large{}Renato M. Fonseca}\thanks{E-mail: renato.fonseca@ific.uv.es} \date{}}

\maketitle

\lyxaddress{\begin{center}
{\Large{}\vspace{-0.5cm}}AHEP Group, Instituto de F\'isica Corpuscular,
C.S.I.C./Universitat de Val\`encia\\
Edif\'icio de Institutos de Paterna, Apartado 22085, E--46071 Val\`encia,
Spain
\par\end{center}}

\begin{center}
23 June 2015
\par\end{center}
\begin{abstract}
The Standard Model (SM) is a chiral theory, where right- and left-handed
fermion fields transform differently under the gauge group. Extra
fermions, if they do exist, need to be heavy otherwise they would
have already been observed. With no complex mechanisms at work, such
as confining interactions or extra-dimensions, this can only be achieved
if every extra right-handed fermion comes paired with a left-handed
one transforming in the same way under the Standard Model gauge group,
otherwise the new states would only get a mass after electroweak symmetry
breaking, which would necessarily be small ($\sim100\textrm{ GeV}$).
Such a simple requirement severely constrains the fermion content
of Grand Unified Theories (GUTs). It is known for example that three
copies of the representations $\mathbf{\overline{5}}+\mathbf{10}$
of $SU(5)$ or three copies of the $\mathbf{16}$ of $SO(10)$ can
reproduce the Standard Model's chirality, but how unique are these
arrangements? In a systematic way, this paper looks at the possibility
of having non-standard mixtures of fermion GUT representations yielding
the correct Standard Model chirality. Family unification is possible
with large special unitary groups --- for example, the $\mathbf{171}$
representation of $SU(19)$ may decompose as $3\left(\mathbf{16}\right)+\mathbf{120}+3\left(\mathbf{1}\right)$
under $SO(10)$.
\\
\\
\\
\\
\\
\\
\\
\\

\end{abstract}

\newpage{}

\section{Introduction}

There is currently no explanation for the flavor structure of the
Standard Model (SM) and Grand Unified Theories (GUTs) developed over
the past decades have failed to shed light on this issue since particles
with different flavors are usually assigned to distinct copies of
a single representation of the enlarged gauge group. For example,
with $SU(5)$ one considers three copies (one for each flavor) of
the representations $\mathbf{\overline{5}}$ and $\mathbf{10}$, containing
exactly the SM fermions \cite{Georgi:1974sy}: three replicas of $Q=\left(\mathbf{3},\mathbf{2},\frac{1}{6}\right)$,
$u^{c}=\left(\overline{\mathbf{3}},\mathbf{1},-\frac{2}{3}\right)$,
$d^{c}=\left(\overline{\mathbf{3}},\mathbf{1},\frac{1}{3}\right)$,
$L=\left(\mathbf{1},\mathbf{2},-\frac{1}{2}\right)$, and $e^{c}=\left(\mathbf{1},\mathbf{1},1\right)$.
In $SO(10)$ models, three $\mathbf{16}$'s contain all SM fermions
plus three right-handed neutrinos $N^{c}=\left(\mathbf{1},\mathbf{1},0\right)$
\cite{Georgi:1974my,Fritzsch:1974nn}. Once the $SO(10)$ symmetry
is broken, a vector (Majorana) mass $mN^{c}N^{c}$ is allowed for
each of these extra fermion states, explaining why they have yet to
be (directly) observed. Increasing further the size of the group,
there is also the well known possibility of having three copies of
the $\mathbf{27}$ in $E_{6}$-based models \cite{Gursey:1975ki},
which contain 11 additional vector particles per generation.

In order to completely explain flavor with GUTs it would be necessary
to place the SM fermions in a single representation of the gauge group.\footnote{Rigorously speaking, when referring to a group we have in mind its
algebra.} This idea goes by the name of \textit{family unification} and it
was attempted in the past with a variety of groups \cite{GellMann:1980vs,Wilczek:1981iz,Fujimoto:1981bv,Chang:1985jd,Bagger:1984rk,Hubsch:1985zn,Bagger:1985mf,Chang:1985uf,Bagger:1984gz,Senjanovic:1984rw,Ida:1980ea,Masiero:1982xr,Sato:1981ga,Sato:1980jn}.
For instance, the spinor representation of $SO(10+2N')$ can be broken
into $2^{N'-1}$ copies of the $\mathbf{16}$ of $SO(10)$, yet it
also contains an equal amount of $\overline{\mathbf{16}}$'s. Therefore,
without confining interactions \cite{GellMann:1980vs,Wilczek:1981iz,Adler:2002yg,Adler:2014pga},
extra dimensions \cite{Dixon:1985jw,Greene:1986bm,Greene:1986jb,Ibanez:1987sn,Ibanez:1986tp,Babu:2002ti,Hwang:2002hg,Choi:2002fn,Gogoladze:2003ci,Gogoladze:2003yw,Han:2004qd,Kawamura:2007cm,Kawamura:2009gr,Choi:2010gx,Goto:2013jma},
or some other elaborate mechanism, one cannot give a big mass to all
these mirror families without making all the families super-heavy
as well.\footnote{The presence of confining gauge interactions could be an elegant solution
to this problem, as pointed out in \cite{GellMann:1980vs,Wilczek:1981iz}.
For example, if $SO(18)$ breaks into $SO(10)\times SO(5)$ such that
$\mathbf{256}\rightarrow3\left(\mathbf{16},\mathbf{1}\right)+\left(\mathbf{16},\mathbf{5}\right)+2\left(\overline{\mathbf{16}},\mathbf{4}\right)$
and if $SO(5)$ becomes non-perturbative at some high-scale, then
one would expect that the only fermions which would remain light would
be the three $\mathbf{16}$'s which are $SO(5)$ singlets. However,
it seems difficult to drive $SO(5)$ into a non-perturbative regime
at high energies.}

In fact, mirror families are just part of a larger problem: in general
it is necessary to justify why all types of exotic fermions are heavy.
Take as an example the representation $\mathbf{560}$ of $SO(10)$,
which is the smallest one containing all SM fermions. On top of the
fact that the excess of fermions ($5Q+4u^{c}+4d^{c}+3L+3e^{c}$) over
mirror fermions ($1\overline{Q}+1\overline{u^{c}}+1\overline{d^{c}}+1\overline{L}+1\overline{e^{c}}$)
is not the correct one, there are also fermions in exotic SM representations
such as $\left(\overline{\mathbf{15}},\mathbf{1},\frac{1}{3}\right)$,
and none in matching conjugate representations. Such states could
only acquire an electroweak (EW) scale mass and therefore would have
already been seen at the Tevatron and the LHC.

Perhaps the idea of unifying the three families in a single GUT fermion
representation is too ambitious. One should then also consider models
where the observed fermion states are distributed over various GUT
representations \cite{Frampton:1979cw,Georgi:1979md,Baaklini:1980gd,Frampton:1979tj,Kim:1980ci,Kashibayashi:1982nu,Frampton:1989fu,Chkareuli:1992kd,Berezhiani:1995dt,Chkareuli:2000bm,Shafi:2001iu,Barr:2008pn,Dent:2009pd,CarcamoHernandez:2010im,Albright:2012zt,Byakti:2013uya}.
Such models might still be quite interesting: if the GUT representations
are not just mere copies of one another, the gauge symmetry alone
might explain the existence of non-trivial flavor structures at low energies.

GUTs with an exotic fermion content may also have unusual features
which go against what is usually taken for granted. One of them is
non-standard normalizations of the hypercharge operator, which we
shall now discuss. In order to see if the gauge couplings unify at
a high scale in a given model, one usually takes the values of $g_{1}=\sqrt{5/3}g'$
, $g_{2}=g$ and $g_{3}=g_{s}$ at roughly the Z-boson mass scale
and runs them with the renormalization group equations up to high
energies. The explanation for the numerical factor $\sqrt{5/3}$ is
simple: the Lagrangian depends on the product of the gauge coupling
constant $g'$ times the hypercharge operator $Y$, so the change
$\left(g',Y\right)\rightarrow\left(n^{-1}g',nY\right)$ for some $n$
is of no consequence in the SM. Comparing the three gauge coupling
constants is then pointless. However, if $SU(3)_{C}\times SU(2)_{L}\times U(1)_{Y}$
is a remnant of a larger simple gauge group, then suddenly there is
a natural value for this $n$ parameter: one would like $Y$ to be
one of the generators of this enlarged group, in which case its normalization
must be the same as the rest of the generators: $\textrm{Tr}\left(T_{a}^{2}\right)=\textrm{constant}$.
As such, if we identify the components of the $\overline{\boldsymbol{5}}$
in an $SU(5)$ theory with those of the SM representations $d^{c}=\left(\overline{\mathbf{3}},\mathbf{1},\frac{1}{3}n\right)$
and $L=\left(\mathbf{1},\mathbf{2},-\frac{1}{2}n\right)$, then $Y=n\times\textrm{diag}\left(\frac{1}{3},\frac{1}{3},\frac{1}{3},-\frac{1}{2},-\frac{1}{2}\right)$
and the third generator of $SU(2)_{L}$ is given by the matrix $T_{3L}=\textrm{diag}\left(0,0,0,\frac{1}{2},-\frac{1}{2}\right)$,
hence $\left|n\right|=\sqrt{3/5}$.

This hypercharge normalization factor is often mentioned as being
specific to $SU(5)$ models, even though it is actually very generic.
Nevertheless, it is conceivable that $n$ might differ from $\sqrt{3/5}$:
for instance, in $SU(5)$ one might try to identify $d^{c}$ with
the representation $\left(\overline{\mathbf{3}},\mathbf{1},-\frac{2}{3}\sqrt{3/5}\right)$
inside the $\boldsymbol{10}$, $u^{c}$ with $\left(\overline{\mathbf{3}},\mathbf{1},\frac{4}{3}\sqrt{3/5}\right)$
in the $\boldsymbol{45}$, and so on, in which case $g_{1}=-\frac{1}{2}\sqrt{5/3}g'$
would be the correct relation.\footnote{As explained later on, this particular example fails because one
would not find anywhere the representation $\left(\mathbf{3},\mathbf{2},-\frac{1}{3}\sqrt{3/5}\right)$
needed for the left-handed quarks.} Another possibility would be to have, for example, an $SU(7)$ model
with the fundamental representation breaking into $X\equiv\left(\overline{\mathbf{3}},\mathbf{1},\frac{1}{3}n\right)+\left(\mathbf{1},\mathbf{2},-\frac{1}{2}n\right)+\left(\mathbf{1},\mathbf{1},m\right)+\left(\mathbf{1},\mathbf{1},-m\right)$
with a non-zero $m$, in which case it is clear that $n$ will not
be equal to $\sqrt{3/5}$, so $g_{1}\neq\sqrt{5/3}g'$ if we were
to identify $d^{c}$ and $L$ with the first two SM representations.
Note that $X$ forms a unitary 7-dimensional (reducible) representation
of the SM group, so it is certainly possible to make the fundamental
representation of $SU(7)$ break in this way.

What about having $\boldsymbol{7}\rightarrow\left(\overline{\mathbf{3}},\mathbf{1},\frac{1}{3}n\right)+\left(\mathbf{1},\mathbf{2},-\frac{1}{2}n\right)+\left(\mathbf{1},\mathbf{2},0\right)$?
Such a scenario is even more interesting. The fundamental representation
of $SU(7)$ branches into a single color anti-triplet, as usual, but
now there are two doublets, which means that one should take $1/\left(2\sqrt{2}\right)$
times the Pauli matrices as the generators of $SU(2)_{L}$, and not
half the Pauli matrices. In such a model, the correctly normalized
gauge couplings would then be $g_{1}=\sqrt{5/3}g'$ , $g_{2}=\sqrt{2}g$
and $g_{3}=g_{s}$.\footnote{It is amusing to consider the possibility of (almost) unifying the
three gauge couplings at low energies exclusively in this way (although
baryon number violation would be a concern). Conceptually, it is not
very complicated: for example, if the fundamental representation of
$SU(9)$ is broken into $\left(\overline{\mathbf{3}},\mathbf{1},\frac{1}{3}n\right)+\left(\mathbf{1},\mathbf{2},-\frac{1}{2}n\right)+\left(\mathbf{1},\mathbf{2},n\right)+\left(\mathbf{1},\mathbf{2},-n\right)$
with $n$ necessarily equal to $\sqrt{3/29}$, successfully associating
the $d^{c}$ and $L$ fermions of the SM with the first two representations
would imply that $g_{1}$, $g_{2}$ and $g_{3}$ have almost the same
value at the EW scale (up to around $\sim10\%$).} So, despite the widespread belief that this issue only affects abelian
groups, clearly there might be potentially interesting normalization
corrections to any of the gauge coupling constants in GUTs with non-standard
fermion assignments.\\

In summary, with or without family unification, it seems appropriate
to systematically study the possible ways of arranging the fermions
in Grand Unified Theories. The requirement that the SM chirality must
be reproduced is a simple yet very stringent constraint which can
be readily used to narrow down the list of possibilities. The aim
of this paper is precisely to analyze, in a comprehensive way, the
fermion sector of GUTs based on different groups, checking whether
or not it is possible to obtain only the observed three families of
fermions plus vector particles. Importantly and in contrast to what
is almost universally done in the literature, the fact that the SM
group can be embedded in more than one way in a given GUT group will
not be overlooked. The aim of the present work is therefore somewhat
similar to the one of the papers \cite{GellMann:1976pg,Georgi:1979md,Frampton:1979tj,King:1981bg},
but it is substantially broader in scope. For example, comparing with
the interesting paper \cite{King:1981bg}, we do not require (a) asymptotic
freedom of the GUT (which severely restricts the number of fermion
components allowed and consequently the group), nor (b) absence of
gauge anomalies (although they get canceled automatically in almost
all cases) and, above all, we do not make the (c) ``bold assumption''
that the embedding of the SM group is as trivial as possible. 

We shall first provide some generic considerations about the method
used to scan over the various GUT groups, representations and embeddings
(section \ref{sec:Framework}) and, following that, the results for
each group are presented and discussed (section \ref{sec:Combinations-of-GUT}
supplemented by an \hyperref[sec:Appendix-A]{appendix}). The main
conclusions are summarized at the end (section \ref{sec:Discussion-and-concluding}).

\section{Framework of the analysis\label{sec:Framework}}

\subsection{The chirality of GUT models and representations}

Let us briefly discuss and settle on a precise definition of (SM)
chirality. Consider some embedding of the SM group $G_{SM}$ in a
bigger group $G$. We shall be interested in tracking the representations
$R_{SM}^{i}$ of $G_{SM}$ contained in some fermion representation
$R$ of $G$ --- the so-called \textit{branching rules} of $R$. Yet,
since pairs of SM vector fermions are irrelevant for the present analysis
(as they can be made very heavy), we may define the chirality of $R$
to be the vector $\chi\left(R\right)$ with component $i$ given by
the number of SM representations $R_{SM}^{i}$ contained in $R$ minus
the number of SM representations $R_{SM}^{i*}$ in $R$:
\begin{align}
\chi_{i}\left(R\right) & \equiv\left(\#R_{SM}^{i}\in R\right)-\left(\#R_{SM}^{i*}\in R\right)\,.
\end{align}
For any real\footnote{In this work, real representations are those which are equivalent
to their conjugate, so they include what is sometimes called real
and pseudo-real representations in other contexts.} SM representation ($R_{SM}^{i}=R_{SM}^{i*}$) we always get $\chi_{i}\left(R\right)=0$.
On the other hand, we have the relation $\chi\left(R^{*}\right)=-\chi\left(R\right)$
which implies that $\chi\left(R\right)$ is the null vector for a
real $R$ ($R=R^{*}$). As such, SM (or GUT) real representations can
be ignored completely and furthermore, concerning complex representations,
the effect of having $n$ copies of $R_{SM}^{i}$(or $R$) in a model
is the same as subtracting from it $n$ copies of $R_{SM}^{i*}$(or
$R^{*}$) as far as chirality is concerned. For this reason, in this
work we take $-R_{SM}^{i}$(or $-R$) to be the exactly the same as
$R_{SM}^{i*}$(or $R^{*}$).

In the case of sums of representations of $G$, chirality is taken
to be simply the sum of the chirality of each representation,
\begin{align}
\chi_{i}\left(R^{1}+\cdots+R^{n}\right) & \equiv\chi_{i}\left(R^{1}\right)+\cdots+\chi_{i}\left(R^{n}\right)\,,
\end{align}
so we can speak of the chirality of a model. For example, in the basis
where we consider only $R_{SM}^{i}=Q,u^{c},d^{c},L,e^{c}$ the chirality
of the Standard Model is given by the vector $\chi\left(SM\right)=\left(3,3,3,3,3\right)^{T}$.\\

This definition of chirality encodes in a precise way the intuitive
notion associated to this word. It counts the number of each type
of SM representation, factoring out real and conjugate pairs of representations.
In the following, we shall see how it allows us to turn the problem
of finding GUTs with the SM chirality into solving a system of linear
equations.

\subsection{GUTs with the correct chirality}

With the above definition, finding the chirality of a representation
of a group $G\supset G_{SM}$ is a matter of decomposing it into SM
representations. In order to do so, one must first know how $G_{SM}$
is embedded in $G$, and it turns out that figuring all the possible
ways of doing so is a complicated problem, which we shall discuss
later. For now, we may assume that this embedding information is
known and fixed. If so, one can use computer programs such as \texttt{Susyno}
\cite{Fonseca:2011sy}\footnote{\label{fn:Notation_comment}The naming scheme for representations
used by the program and in this work follows the convention laid out
in the manual of the program \texttt{LieART} \cite{Feger:2012bs}.} or \texttt{LieART} \cite{Feger:2012bs} to decompose in a systematic
way any representation of the group $G$ into those of $G_{SM}$ (for
this work, the former was used).

In turn, with the branching rules of a list of representations $R_{i}$
of $G$, it is a rather simple exercise of linear algebra to find
all integer linear combinations $\sum_{i}c_{i}R_{i}$ of the $R_{i}$
with the SM chirality. Indeed, defining $\mathcal{M}$ to be the matrix
with entries $\mathcal{M}_{ij}=\chi_{i}\left(\Psi_{j}\right)$, the
vector $c=\left(c_{1},\cdots,c_{n}\right)^{T}$ whose components are
the integers $c_{i}$ we seek is the solution to the linear system
\begin{align}
\chi\left(SM\right) & =\mathcal{M}\cdot c\label{eq:main_equation}
\end{align}
where $\chi\left(SM\right)$ is a vector with the SM chirality,
as mentioned previously. From $\chi\left(SM\right)$ and $\mathcal{M}$,
one can extract $c$. As it is well known, the general solution of
this equation is of the form
\begin{align}
c & =\widetilde{c}+\sum_{i}\alpha_{i}n_{i}\,,\label{eq:main_solution}
\end{align}
where $\widetilde{c}$ is any particular solution of equation \eqref{eq:main_equation}
and the vectors $n_{i}$ are a basis of the nullspace of $\mathcal{M}$
(i.e.,$\mathcal{M}\cdot n_{i}=0$). The $\alpha_{i}$ are plain numbers
which can take any value, as long as the components of the $c$ vector
are integer numbers. One can understand this generic form of $c$
as follows: the vector $\widetilde{c}$ describes a particular combination
of the $R_{j}$ ($\sum_{j}\widetilde{c}_{j}R_{j}$) possessing the
correct chirality, and each of the $n_{i}$ describes an independent,
non-trivial combination of the $R_{j}$ ($\sum_{j}\left(n_{i}\right)_{j}R_{j}$)
with no chirality; therefore an arbitrary number of $n_{i}$'s can
be added or subtracted to $\widetilde{c}$.\\

To clarify this approach, consider the following straightforward example.
Take $SU(5)$ as the grand unified group, and its complex representations
up to size 35 (we only need to consider one member of each conjugated
pair): $\mathbf{5},\mathbf{10},\mathbf{15},\mathbf{35}$. They decompose
into the following eleven $SU(3)_{C}\times SU(2)_{L}\times U(1)_{Y}$
representations (plus their conjugates): $\left(\mathbf{1},\mathbf{1},1\right)$,
$\left(\mathbf{1},\mathbf{2},-\frac{1}{2}\right)$, $\left(\mathbf{1},\mathbf{3},-1\right)$,
$\left(\mathbf{1},\mathbf{4},-\frac{3}{2}\right)$, $\left(\overline{\mathbf{3}},\mathbf{1},\frac{1}{3}\right)$,
$\left(\mathbf{3},\mathbf{2},\frac{1}{6}\right)$, $\left(\overline{\mathbf{3}},\mathbf{3},-\frac{2}{3}\right)$,
$\left(\mathbf{6},\mathbf{1},\frac{2}{3}\right)$, $\left(\mathbf{6},\mathbf{2},\frac{1}{6}\right)$,
$\left(\overline{\mathbf{10}},\mathbf{1},1\right)$, $\left(\overline{\mathbf{3}},\mathbf{1},-\frac{2}{3}\right)$.
With this ordering of the SM representations, from the decomposition
of $R=\mathbf{5},\mathbf{10},\mathbf{15},\mathbf{35}$ we get
\begin{align}
\chi\left(\mathbf{5}\right) & =\left(0,-1,0,0,-1,0,0,0,0,0,0\right)^{T}\,;\quad\chi\left(\mathbf{10}\right)=\left(1,0,0,0,0,1,0,0,0,0,1\right)^{T}\,,\\
\chi\left(\mathbf{15}\right) & =\left(0,0,-1,0,0,1,0,-1,0,0,0\right)^{T}\,,\quad\chi\left(\mathbf{35}\right)=\left(0,0,0,1,0,0,1,0,1,1,0\right)^{T}\,.
\end{align}
The chirality of the SM itself is
\begin{align}
\chi\left(SM\right) & =\left(3,3,0,0,3,3,0,0,0,0,3\right)^{T}\,.
\end{align}
So, how many copies of each of the four $SU(5)$ representations are
needed in order to obtain the SM chirality? If the fermions of the
GUT model are $c_{1}\left(\mathbf{5}\right)+c_{2}\left(\mathbf{10}\right)+c_{3}\left(\mathbf{15}\right)+c_{4}\left(\mathbf{35}\right)$,
solving the system
\begin{align}
\left(\begin{array}{c}
3\\
3\\
0\\
0\\
3\\
3\\
0\\
0\\
0\\
0\\
3
\end{array}\right) & =\left(\begin{array}{cccc}
0 & 1 & 0 & 0\\
-1 & 0 & 0 & 0\\
0 & 0 & -1 & 0\\
0 & 0 & 0 & 1\\
-1 & 0 & 0 & 0\\
0 & 1 & 1 & 0\\
0 & 0 & 0 & 1\\
0 & 0 & -1 & 0\\
0 & 0 & 0 & 1\\
0 & 0 & 0 & 1\\
0 & 1 & 0 & 0
\end{array}\right)\cdot\left(\begin{array}{c}
c_{1}\\
c_{2}\\
c_{3}\\
c_{4}
\end{array}\right)
\end{align}
yields a unique solution (the matrix in this equation has a trivial
nullspace): $\left(c_{1},c_{2},c_{3},c_{4}\right)=\left(-3,3,0,0\right)$.
This means that, with $SU(5)$ representations of size up to 35,
the fermion fields must be $-3\left(\boldsymbol{5}\right)+3\left(\mathbf{10}\right)$
or equivalently $3\left(\overline{\boldsymbol{5}}\right)+3\left(\mathbf{10}\right)$,
up to trivial variations (i.e., addition of real representations or
conjugate pairs of complex ones).

Needless to say, the conclusion reached with this example is unremarkable
given that the representations considered were just $R=\mathbf{5},\mathbf{10},\mathbf{15},\mathbf{35}$.
The analysis gets more interesting when bigger representations are
considered. If we do so, how unique is the standard fermion content
$3\left(\overline{\boldsymbol{5}}\right)+3\left(\mathbf{10}\right)$
in $SU(5)$ GUTs? This is an important question which we address
in this work, noting that the normalization of the SM hypercharge
(usually given by a factor $\sqrt{3/5}$) depends on its answer.

Unfortunately, the type of simple analysis just presented is complicated
by the fact that the SM group may be embedded in a GUT group $G$
in more than one way. In particular, it is not known a priori what
are the valid ways of combining the multiple $U(1)$ factors inside
$G$ in order to form the SM's $U(1)_{Y}$.

\subsection{\label{sub:Different-ways-of}Different ways of embedding $G_{SM}$
in a group $G\supset G_{SM}$}

A systematic study of the different ways in which the SM chirality
can be achieved in a GUT based on a group $G$ must necessarily take
into account the distinct ways in which $G_{SM}$ can be embedded
in $G$. (In fact, we only need to care about branching rules, so
we shall be pragmatic and equate \textit{different embeddings} to
\textit{different branching rules}.) Regardless of the actual symmetry
breaking chain, we can can view it as being made of two symbolic steps:
\begin{equation}
G\overset{(1)}{\rightarrow}SU(3)_{C}\times SU(2)_{L}\times U\left(1\right)^{m}\overset{(2)}{\rightarrow}SU(3)_{C}\times SU(2)_{L}\times U\left(1\right)_{Y}\,.
\end{equation}
In the first step, $G$ is reduced to $SU(3)_{C}\times SU(2)_{L}$
times a maximal number of $U\left(1\right)$ factors, while in a second
step this abelian part of the group is reduced to $U\left(1\right)_{Y}$.

There is only a finite number of ways in which the first symmetry
breaking step can be carried out (see table \ref{tab:NumberOfGSMEmbbedings}).
Indeed, with the information in \cite{Dynkin:1957_1,Dynkin:1957_2}
one can break any semi-simple Lie algebra in a step-wise manner, $G\rightarrow G'\rightarrow G''\rightarrow\cdots\rightarrow SU(3)_{C}\times SU(2)_{L}\times U\left(1\right)^{m}$,
such that the algebra of each group in this sequence is a maximal
subalgebra of the preceding one,\footnote{It is said that $G'$ is a maximal subalgebra of $G$ if there is
no subalgebra $G''$ of $G$ such that $G'\subset G''\subset G$ (other
than the trivial cases $G''=G$ or $G'$).} discarding none of the $U\left(1\right)$ factors at this stage.

There are two important points concerning this first symmetry breaking
step. The first one is that the number of $U\left(1\right)$ factors
in the end result ($SU(3)_{C}\times SU(2)_{L}\times U\left(1\right)^{m}$)
will depend on the chosen sequence of maximal subalgebras, as the
rank of the groups may shrink. The second point is that the step-wise
procedure of breaking $G$ into $SU(3)_{C}\times SU(2)_{L}\times U\left(1\right)^{m}$
subgroups will in general produce a large number of repetitions. In
order to verify whether two embeddings are indeed different, it suffices
to check that the branching rules for the fundamental representation
of $G$ are distinct, with the exception of the $SO\left(2n\right)$
groups which also require the branching rules of the spinor representation
to be distinct \cite{Dynkin:1957_2,Losev:2010,Minchenko:2006}.

To illustrate these two remarks, consider the case of $G=SO\left(10\right)$.
Its maximal subgroups are $SU\left(5\right)\times U(1)$, $SU(4)\times SU(2)\times SU(2)$,
$SO(8)\times U(1)$, $SP\left(4\right)$, $SO(9)$, $SU(2)\times SO(7)$,
$SP(4)\times SP(4)$ although only the first two correspond to chiral
embeddings.\footnote{By chiral embeddings we are referring to those cases $G\supset G'$
where the representations of $G$ do not break only into real representations
and conjugate pairs of complex representations of $G'$.} In any case, consider for example the breaking chains (A) $SO(10)\rightarrow SU\left(5\right)\times U(1)\rightarrow SU(3)\times SU(2)\times U\left(1\right)^{2}$,
(B) $SO(10)\rightarrow SU(4)\times SU(2)\times SU(2)\rightarrow SU(3)\times SU(2)\times U\left(1\right)^{2}$,
and (C) $SO(10)\rightarrow SU(2)\times SO(7)\rightarrow SU(3)\times SU(2)\times U\left(1\right)$.
The fundamental and spinor representations of $SO(10)$ branch as
follows,\thinmuskip=0.5mu
\medmuskip=1mu 
\thickmuskip=1.5mu
\begin{align}
\left(\begin{array}{@{}c@{}}
\mathbf{10}\\
\mathbf{16}
\end{array}\right) & \overset{A,B}{\rightarrow}\left(\begin{array}{@{}c@{}}
\left(\mathbf{3},\mathbf{1},0,2\right)+\left(\overline{\mathbf{3}},\mathbf{1},0,-2\right)+\left(\mathbf{1},\mathbf{2},1,0\right)+\left(\mathbf{1},\mathbf{2},-1,0\right)\\
\left(\mathbf{3},\mathbf{2},0,-1\right)+\left(\overline{\mathbf{3}},\mathbf{1},1,1\right)+\left(\overline{\mathbf{3}},\mathbf{1},-1,1\right)+\left(\mathbf{1},\mathbf{2},0,3\right)+\left(\mathbf{1},\mathbf{1},-1,-3\right)+\left(\mathbf{1},\mathbf{1},1,-3\right)
\end{array}\right)\nonumber \\
 & \,\textrm{ or }\left(\begin{array}{@{}c@{}}
\left(\mathbf{3},\mathbf{1},0,2\right)+\left(\overline{\mathbf{3}},\mathbf{1},0,-2\right)+\left(\mathbf{1},\mathbf{2},1,0\right)+\left(\mathbf{1},\mathbf{2},-1,0\right)\\
\left(\overline{\mathbf{3}},\mathbf{2},0,-1\right)+\left(\mathbf{3},\mathbf{1},1,1\right)+\left(\mathbf{3},\mathbf{1},-1,1\right)+\left(\mathbf{1},\mathbf{2},0,3\right)+\left(\mathbf{1},\mathbf{1},-1,-3\right)+\left(\mathbf{1},\mathbf{1},1,-3\right)
\end{array}\right)\,,\label{eq:10}\\
\left(\begin{array}{@{}c@{}}
\mathbf{10}\\
\mathbf{16}
\end{array}\right) & \overset{C}{\rightarrow}\left(\begin{array}{@{}c@{}}
\left(\mathbf{3},\mathbf{1},2\right)+\left(\overline{\mathbf{3}},\mathbf{1},-2\right)+\left(\mathbf{1},\mathbf{3},0\right)+\left(\mathbf{1},\mathbf{1},0\right)\\
\left(\mathbf{3},\mathbf{2},-1\right)+\left(\overline{\mathbf{3}},\mathbf{2},1\right)+\left(\mathbf{1},\mathbf{2},3\right)+\left(\mathbf{1},\mathbf{2},-3\right)
\end{array}\right)\,,\label{eq:11}
\end{align}
\thinmuskip=1mu
\medmuskip=4mu plus 2mu minus 4mu
\thickmuskip=5mu plus 5muusing unnormalized $U(1)$ charges (the separation of the two $U\left(1\right)$'s
is irrelevant; the $U\left(1\right)^{2}$ group as a whole should
be the same). Paths (A) and (B) lead to the same branching rules:
there are two of them, which are related to one-another by conjugation
of the color quantum number. This is a trivial variation which exists
for all chiral embeddings, so we may refer to `pairs of chiral embeddings',
although in the next section we shall simply focus on one member of
each such pair of embeddings.

So overall we could say that there are three possible embeddings of
$G_{SM}$ in $SO(10)$ (having in mind step one of the symmetry breaking
only), including one pair which is chiral. In the second symmetry
breaking step, the two $U\left(1\right)$'s can be combined in any
way to form $U(1)_{Y}$ --- at least from a purely group theoretical
perspective. Therefore, the branching rules of the natural and spinor
representations under the symmetry breaking $SO(10)\rightarrow G_{SM}$
are (the normalization of the $U(1)$'s is irrelevant at the moment)\thinmuskip=0.5mu
\medmuskip=1mu 
\thickmuskip=1.5mu 
\begin{align}
\left(\begin{array}{@{}c@{}}
\mathbf{10}\\
\mathbf{16}
\end{array}\right) & \rightarrow\left(\begin{array}{@{}c@{}}
\left(\mathbf{3},\mathbf{1},2\beta\right)+\left(\overline{\mathbf{3}},\mathbf{1},-2\beta\right)+\left(\mathbf{1},\mathbf{2},\alpha\right)+\left(\mathbf{1},\mathbf{2},-\alpha\right)\\
\left(\mathbf{3},\mathbf{2},-\beta\right)+\left(\overline{\mathbf{3}},\mathbf{1},\alpha+\beta\right)+\left(\overline{\mathbf{3}},\mathbf{1},-\alpha+\beta\right)+\left(\mathbf{1},\mathbf{2},3\beta\right)+\left(\mathbf{1},\mathbf{1},-\alpha-3\beta\right)+\left(\mathbf{1},\mathbf{1},\alpha-3\beta\right)
\end{array}\right)\,,\label{eq:SO10_example_emb1}\\
 & \rightarrow\left(\begin{array}{@{}c@{}}
\left(\mathbf{3},\mathbf{1},2\beta\right)+\left(\overline{\mathbf{3}},\mathbf{1},-2\beta\right)+\left(\mathbf{1},\mathbf{2},\alpha\right)+\left(\mathbf{1},\mathbf{2},-\alpha\right)\\
\left(\overline{\mathbf{3}},\mathbf{2},-\beta\right)+\left(\mathbf{3},\mathbf{1},\alpha+\beta\right)+\left(\mathbf{3},\mathbf{1},-\alpha+\beta\right)+\left(\mathbf{1},\mathbf{2},3\beta\right)+\left(\mathbf{1},\mathbf{1},-\alpha-3\beta\right)+\left(\mathbf{1},\mathbf{1},\alpha-3\beta\right)
\end{array}\right)\,,\label{eq:SO10_example_emb2}\\
 & \rightarrow\left(\begin{array}{@{}c@{}}
\left(\mathbf{3},\mathbf{1},2\gamma\right)+\left(\overline{\mathbf{3}},\mathbf{1},-2\gamma\right)+\left(\mathbf{1},\mathbf{3},0\right)+\left(\mathbf{1},\mathbf{1},0\right)\\
\left(\mathbf{3},\mathbf{2},-\gamma\right)+\left(\overline{\mathbf{3}},\mathbf{2},\gamma\right)+\left(\mathbf{1},\mathbf{2},3\gamma\right)+\left(\mathbf{1},\mathbf{2},-3\gamma\right)
\end{array}\right)\,,\label{eq:SO10_example_emb3}
\end{align}
\thinmuskip=1mu
\medmuskip=4mu plus 2mu minus 4mu
\thickmuskip=5mu plus 5mufor some $\alpha,\beta,\gamma$ factors. It is worth stressing that
the last case is not a special case of either of the first two. Also,
while we did only consider three maximal subgroups of $SO(10)$ (chains
A, B, C) it can be checked that all others cases lead to the embedding
of the form \eqref{eq:SO10_example_emb3} --- doing so by hand is
tedious (and even more so for bigger groups), and for that reason
the \texttt{Susyno} program was used to automatically check for these
repetitions.

Introducing physical considerations, not all embeddings of forms in
equations \eqref{eq:SO10_example_emb1}--\eqref{eq:SO10_example_emb3}
can be used to embed the SM in an $SO(10)$ GUT. The one in equation
\eqref{eq:SO10_example_emb3} would lead to a vector theory, so it
can be excluded. As for the chiral embedding described by equation
\eqref{eq:SO10_example_emb1} (\eqref{eq:SO10_example_emb2} is similar),
if we are to obtain the SM fermions from the $\mathbf{16}$, then
$\alpha$ and $\beta$ which describe the composition of $U(1)_{Y}$
in terms of the two $U(1)$'s of $SU(3)_{C}\times SU(2)_{L}\times U\left(1\right)^{2}$
must take specific values: $\beta=-1/6$ and $\alpha=\pm1/2$. This
leads to the standard $G_{SM}$ embedding in $SO(10)$, yet, as it
should be clear at this point, $U(1)_{Y}$ might conceivably be another
combinations of the two $U(1)$'s if the SM fermions are placed in
other $SO(10)$ representations. The normalization of the hypercharge
may depend on this placement.

Unfortunately, with larger GUT groups the situation becomes even more
complicated if we do not make assumptions about the GUT representations
where the SM fields are embedded, since there are more $U(1)$ factors
to consider. This situation is not insurmountable, but it does require
adaptations to the analysis suggested in section \ref{sec:Framework},
since it cannot be carried out unless we know the hypercharge $y$
of the representations (all that is known is that $y=\sum_{i}\alpha_{i}y_{i}$
where $y_{i}$ are the charges under $U\left(1\right)^{m}$, and the
$\alpha_{i}$ are to be determined).

There seems to be no easy way to tackle this issue, and as a consequence
the scans over GUT representations were smaller for the bigger groups.
One way is to just look at the first two quantum numbers and try to
match in all possible ways the SM representations with the ones of
$SU(3)_{C}\times SU(2)_{L}\times U\left(1\right)^{m}$ obtained from
some list of GUT representations --- this is the intuitive approach
which works very well for the $\mathbf{16}$ of $SO(10)$. Whenever
this approach proved to be too demanding computationally, we used
instead a modification of the analysis in section \ref{sec:Framework},
where the $c_{i}$ (encoding the unknown combination of the GUT representations)
and the $\alpha_{i}$ (encoding the unknown combination of the $U(1)$'s)
are found simultaneously as the solution of complicated equations
where the $\alpha_{i}$ do not appear linearly.

\begin{center}
\begin{table}
\begin{centering}
\begin{tabular}{cccccccc}
\toprule 
Group & \# Embeddings &  & Group & \# Embeddings &  & Group & \# Embeddings\tabularnewline
\cmidrule{1-2} \cmidrule{4-5} \cmidrule{7-8} 
$SU\left(5\right)$ & 2 (1) &  & $SU\left(9\right)$ & 40 (19) &  & $SO\left(10\right)$ & 3 (1)\tabularnewline
$SU\left(6\right)$ & 4 (2) &  & $SU\left(10\right)$ & 65 (30) &  & $SO\left(14\right)$ & 15 (2)\tabularnewline
$SU\left(7\right)$ & 10 (5) &  & $SU\left(11\right)$ & 108 (50) &  & $SO\left(18\right)$ & 62 (5)  \tabularnewline
$SU\left(8\right)$ & 21 (10) &  & $SU\left(12\right)$ & 187 (86) &  & $E_{6}$ & 12 (5)\tabularnewline
\bottomrule
\end{tabular}
\par\end{centering}

\protect\caption{\label{tab:NumberOfGSMEmbbedings}Number of distinct embeddings of
subgroups of the type $H=SU(3)\times SU(2)\times U\left(1\right)^{m\geq1}$
in various simple groups $G$ which have complex representations (the
number of pairs of chiral embeddings is indicated in parenthesis).
The counting includes only those cases where $H$ is not contained
in a bigger subgroup $H'=SU(3)\times SU(2)\times U\left(1\right)^{m'}$
with $m'>m$. Note that, for a given $G$, the number of $U(1)$ factors
of the $H$ subgroups in this condition does not need to be constant
(consider the $SO\left(10\right)$ example in the main text).}
\end{table}

\par\end{center}

\section{\label{sec:Combinations-of-GUT}GUTs with the SM chirality}

The GUTs we wish to consider should be based on a group with complex
representations, otherwise they would not give rise to an effective
chiral theory. The simple Lie groups with this property and which
contain $G_{SM}$ as a subgroup are $SU(N\geq5)$, $SO(4N'+2)$ for
$N'\geq2$, and $E_{6}$. As such, in the following we shall analyze
the fermion sector of GUTs based on one of these simple groups, investigating
also models with the $SU(3)\times SU\left(3\right)$ gauge group (possibly
with an extra $U(1)$).

\subsection{$SU\left(5\right)$}

We shall start by assuming that the hypercharge of the SM particles
are normalized in the usual way: $y\left(e^{c}\right)=\sqrt{3/5}$
for example. All 2048 (pairs of) complex representations of $SU(5)$
with size no larger than 1 million were decomposed with the \texttt{Susyno}
program. A total of 29037 SM complex representations appear in these
decompositions, therefore one obtains the system of linear equations
in \eqref{eq:main_equation} where $\chi\left(SM\right)$ is a 29037-dimensional
vector (with null components everywhere except for five $\pm3$ entries),
$c$ is 2048-dimensional vector of unknown coefficients $c_{i}$ (describing
the number of copies of each $SU(5)$ representation), and $\mathcal{M}$
is a 29037 by 2048 matrix. Both $\chi\left(SM\right)$ and $\mathcal{M}$
are known, so it is possible to solve for the vector $c$ as explained
in section \ref{sec:Framework}. It turns out that the matrix $\mathcal{M}$
has a trivial nullspace. As such, there is a single solution to equation
\eqref{eq:main_equation}, and it corresponds to the standard, well
known one: three copies of $\overline{\boldsymbol{5}}+\boldsymbol{10}$.\footnote{We recall here that, since chiral embedding come in pairs (see the
example in equations \eqref{eq:SO10_example_emb1}--\eqref{eq:SO10_example_emb2}),
three copies of $\boldsymbol{5}+\overline{\boldsymbol{10}}$ would
work as well. This is nevertheless an obvious/trivial variation which
we shall ignore in the remainder of this work.}\footnote{Reference \cite{Georgi:1979md} claims the same thing: the hope of
recovering the SM with more exotic $SU\left(5\right)$ representations
is not possible. The author of \cite{Georgi:1979md} supports this
assertion with the fact that the rank of $SU\left(5\right)$ and $G_{SM}$
are the same, however it is not clear exactly how this fact can be
used to proof the statement that the only non-trivial solution with
the SM chirality are three copies of $\overline{\boldsymbol{5}}+\boldsymbol{10}$.

The fact that a group and a subgroup have the same rank implies that
the corresponding weight projection matrix (see \cite{Slansky:1981yr}
for details) is invertible, therefore there is (at most) one $SU\left(5\right)$
field content which can break into a given combination of $G_{SM}$
representations. However, since we do not mind adding vector fermions
to the SM, there is an infinite set of combinations of $G_{SM}$ representations
which are acceptable. Out of these, the computer scan done in the
current analysis shows that, using representations with size no larger
than 1 million, the only valid combination of SM fermions which can
come from an $SU\left(5\right)$ theory is the one associated to three
copies of $\overline{\boldsymbol{5}}+\boldsymbol{10}$ (and trivial
variations of it).

Furthermore, as explained in the main text, one can conceivably take
a SM hypercharge normalization distinct from the canonical one, which
further complicates the use of the above group rank argument to rule
out non-trivial $SU\left(5\right)$ solutions with the SM chirality.}

This simple but effective analysis shows that the $\overline{\boldsymbol{5}}$
and $\boldsymbol{10}$ fermion representations of $SU(5)$ are extremely
special. However, we did assume the standard GUT hypercharge normalization
factor $n_{strd}\equiv\sqrt{\frac{3}{5}}$. The usual justification
for this factor is tied to the identify the components of $\overline{\boldsymbol{5}}$
with those of the SM representations $L=\left(\mathbf{1},\mathbf{2},-\frac{1}{2}n\right)$
and $d^{c}=\left(\overline{\mathbf{3}},\mathbf{1},-\frac{1}{3}n\right)$,
as discussed in the introduction of this document. Since we do not
want to assume that the SM fermions are in the $\overline{\boldsymbol{5}}$
and $\boldsymbol{10}$ representations necessarily, we must admit
other values for $n$. Which other values can it take? Looking for
$SU(5)$ representations where the left-handed quarks might be embedded,
we conclude that the $G_{SM}$ representations of the form $\left(\mathbf{3},\mathbf{2},\frac{1}{6}n\right)$
must have $n=\left(1+6k\right)n_{strd}$ for some integer $k$. This
can be easily shown analytically with the weight projection method
(the reader may wish to see for example \cite{Slansky:1981yr}) and
indeed, probing the $SU(5)$ representations of size smaller or equal
to a million, one encounters all the SM representations fermions $\left\{ Q,u^{c},d^{c},L,e^{c}\right\} $
with the hypercharge normalizations $n/n_{strd}=-17,-11,-5,1,7,13,19$.
Crucially, each of these choices yields a different chirality vector
$\chi\left(SM\right)$ in equation \eqref{eq:main_equation}, and
it turns out that there is no solutions except for the standard hypercharge
normalization ($n/n_{strd}=1$).

\subsection{$SO\left(10\right)$}

We now repeat for $SO\left(10\right)$ the same analysis which was
done for $SU\left(5\right)$. According to the discussion of subsection
\prettyref{sub:Different-ways-of}, there is only one chiral embedding
of $SU(3)\times SU(2)\times U\left(1\right)^{m}$ in $SO\left(10\right)$
which is the one in equation \eqref{eq:SO10_example_emb1} (equation
\eqref{eq:SO10_example_emb2} is similar) yet, since $m=2$, we do
have to probe all possible values of $\alpha$ and $\beta$ which
encode the relation between $U(1)_{Y}$ and the two $U(1)$'s which
are contained in $SO\left(10\right)$. A list of possibilities can
be computed by breaking all $SO\left(10\right)$ representations up
to some size into those of $G'=SU(3)\times SU(2)\times U\left(1\right)^{2}$
and then start assigning the SM fermions (at least two) to any $G'$
representations with the correct $SU(3)\times SU(2)$ quantum numbers.
Such procedure should be compared with the one used for $SU\left(5\right)$
(see above) where, instead of two, there was only one unknown parameter
(a normalization factor).

This method produces an exhaustive list of $\left(\alpha,\beta\right)$
values for which $SO\left(10\right)$ breaks into $SU(3)\times SU(2)\times U(1)$
in such a way that all the SM fermions ($Q$, $u^{c}$, $d^{c}$,
$L$, $e^{c}$) can be found inside some $SO\left(10\right)$ representation,
with the correct ratio of hypercharges. For each value of $\left(\alpha,\beta\right)$
it is then possible to compute an $\mathcal{M}$ matrix and solve
equation \eqref{eq:main_equation} using it. This was done for all
$SO\left(10\right)$ representations up to size 1 million, and two
conclusions became clear.

Firstly, concerning the embedding of $G_{SM}$ in $SO\left(10\right)$
and the normalization of the hypercharge, it turns out that equation
\eqref{eq:main_equation} admits solutions only for the values $\left(\alpha,\beta\right)=\left(\pm\frac{1}{2},-\frac{1}{6}\right)$,
corresponding to
\begin{align}
\mathbf{16} & \rightarrow Q+d^{c}+u^{c}+L+e^{c}+N^{c}\label{eq:SO10_BR}
\end{align}
with the standard hypercharge normalization. It is worth mentioning
that even though $SO\left(10\right)$ contains $SU(5)$, which in
turn contains $G_{SM}=SU(3)\times SU(2)\times U(1)$ as a subgroup,
there are more such $G_{SM}$ subgroups in $SO(10)$, and that is
why the branching rules may vary depending on which one is picked.
With this in mind, it is interesting to note that the only branching
rule which can reproduce the SM chirality (shown above) matches the
one of the $G_{SM}$ subgroup found inside $SU(5)$. We point this
feature now because in the remainder of this work we shall see that
this is not a specific feature of $SO(10)$: for all simple GUT groups
which were tested, in order to recover the SM chirality, the embedding $G_{SM}\subset G_{GUT}$
  must be such that the branching
rules match those for a $G_{SM}$ inside a particular $SU(5)$
subgroup of $G_{GUT}$.\footnote{The branching rules are the same, but this does not mean that $G_{SM}\subset SU\left(5\right)\subset G_{GUT}$
necessarily. Consider the following counter example: $G_{SM}\subset SU\left(3\right)\times SU(2)\times U(1)_{B}\times U(1)_{A}\subset SU\left(5\right)\times U(1)_{A}\subset SO(10)$
and it is possible to form the SM hypercharge group just from the
$U(1)_{A}$ inside $SU(5)$ or from a combination of $U(1)_{A}$ and
$U(1)_{B}$ (known as the flipped $SU(5)$ scenario \cite{Barr:1981qv}).
In either case, the branching rules $SO(10)\rightarrow G_{SM}$ are
the same. Something similar happens with the $E_{6}$ group, as the
hypercharge group of the SM can be made from three different combinations
of the three available $U(1)$'s --- see for instance \cite{Bertolini:2010yz}.}

The second conclusion concerns the valid combinations of $SO\left(10\right)$
fermion fields: unlike $SU(5)$, we do find non-standard solutions
with the SM chirality, although they involve very large representations.
To be precise, referring to the framework set forth in section \ref{sec:Framework},
we have the solution $3\left(\mathbf{16}\right)$ to which we can
add an arbitrary number of the non-trivial combinations of $SO(10)$
representations which have no chirality (associated to the $n_{i}$
vectors of equation \eqref{eq:main_solution}): \thinmuskip=0.5mu
\medmuskip=1mu 
\thickmuskip=1.5mu
\begin{align}
n_{1}: & \;\;\;\;\;\;-\mathbf{126}-\mathbf{144}-\mathbf{1200}+\mathbf{2772}+\mathbf{3696}-\mathbf{4950}-\mathbf{6930'}+\mathbf{7920}+\mathbf{8064}+\mathbf{11088}-\mathbf{15120}\nonumber \\
 & \;\;\;\;\;\;-\mathbf{17280}+\mathbf{17325}+\mathbf{30800}-\mathbf{34992}-\mathbf{38016}+\mathbf{48114}+\mathbf{49280}\,,\\
n_{2}: & \;\;\;\;\;\;-\mathbf{16}+\mathbf{126}-\mathbf{560}+\mathbf{8064}+\mathbf{20592}-\mathbf{20790}+\mathbf{23760}+\mathbf{25200}-\mathbf{29568}-\mathbf{48114}-\mathbf{50050}\nonumber \\
 & \;\;\;\;\;\;-\mathbf{90090}-\mathbf{102960}-\mathbf{124800}-\mathbf{128700}-\mathbf{144144}+\mathbf{164736}+\mathbf{196560}-\mathbf{199017}\,.
\end{align}
\thinmuskip=1mu
\medmuskip=4mu plus 2mu minus 4mu
\thickmuskip=5mu plus 5muWe recall here that $-R$ should be interpreted whenever necessary
as $\overline{R}$. The field combinations $n_{1}$ and $n_{2}$ are
just two out of many with no chirality; there are more such (independent)
mixtures of fields, involving even bigger $SO\left(10\right)$ representations,
which we will not write down here.\footnote{A total of 6 independent $\boldsymbol{n_{i}}$ combinations with no
chirality exist involving only $SO\left(10\right)$ representations
with size smaller or equal to 1 million.} We simply point out here that the $n_{1}$ combination does not involve
the $\mathbf{16}$, so any solution of the form $3\left(\mathbf{16}\right)+kn_{1}$
for some $k\in\mathbb{Z}$ will still involve 3 copies of the spinor
representation. However, $n_{2}$ is a linear combination of representations
which does contain the $\mathbf{16}$: a GUT theory with the fermion
content $3\left(\mathbf{16}\right)+3n_{2}$ would have no fermions
in the spinor representation and still its chirality would be correct.
Therefore, it is possible to build a GUT model based on the $SO\left(10\right)$
group without spinors, even though its matter content would need to
be extremely large\footnote{Quantum gravity effects are expected to become relevant in such a
scenario \cite{Dvali:2007hz,Calmet:2008df}. } and, on the other hand, the flavor problem would persist since 3
copies (or more) of each fermion representation would still be needed.

\subsection{$E_{6}$}

The group $E_{6}$ has 38 pairs of complex representations with size
at most 1 million. There are a total of 12 distinct ways of embedding
$SU(3)\times SU(2)\times U\left(1\right)^{m}$ in $E_{6}$, which
includes 5 pairs of chiral embeddings. For each of these, we have
allowed $U(1)_{Y}$ to be any combination of the $m$ $U\left(1\right)$'s.
Remarkably, once this variety of representations and embeddings is
fully explored, it turns out that there is a unique solution with
the correct chirality. In other words, there is both a unique embedding
and a unique fermion field configuration which yield the SM chirality:
it is 3 copies of the $\boldsymbol{27}$ representation with the embedding
\begin{align}
\boldsymbol{27} & \rightarrow Q+u^{c}+2d^{c}+2L+e^{c}+\left(d^{c}\right)^{*}+L^{*}+2N^{c}\,.
\end{align}
It is well known that this branching rule matches the one for the
$G_{SM}$ subgroup of the $SO\left(10\right)$ which is inside $E_{6}$,
with the $\boldsymbol{27}$ breaking into $\boldsymbol{1}+\boldsymbol{10}+\boldsymbol{16}$
of $SO\left(10\right)$, so it is clear that the hypercharge normalization
factor must be the standard one ($\sqrt{3/5}$).

\subsection{$SU(N>5)$ }

Subsection \ref{sub:Different-ways-of} contains a discussion on how
to obtain all the embeddings of a subgroup $H$ in some group $G$
(with $H=SU(3)\times SU(2)\times U\left(1\right)^{m}$ in mind). It
involves probing all sequences of maximal subgroups $G\rightarrow G'\rightarrow G''\rightarrow\cdots\rightarrow H$,
which can be a very time consuming process that does not provide much
insight on the direct relationship between $H$ and $G$. It is therefore
important to realize that there is a simpler way to list all these
embeddings when $G$ is a special unitary group.

For a given $G=SU(N)$, we start by picking all possible combinations
of the irreducible representations of $H$ in order to obtain a complete
list of the $N$-dimensional (potentially reducible) representations
of the subgroup $H$. Since the representation matrices must be unitary
(otherwise the kinetic terms would not be gauge invariant), it is
obvious that there must be an embedding under which the fundamental
of $G$ breaks into any of the $N$-dimensional (unitary) representations
of $H$. One only needs to ensure that none of the generators of these
$H$ representations has null/trivial generators $T_{a}$ as this
would imply that $\textrm{Tr}\left(T_{a}^{2}\right)=0$.

For example, consider the embedding of $SU(5)$ in $SU(6)$. There
are three inequivalent reducible 6-dimensional representations of
$SU(5)$: $\mathbf{1}+\mathbf{1}+\mathbf{1}+\mathbf{1}+\mathbf{1}+\mathbf{1}$,
$\mathbf{1}+\mathbf{5}$ , and $\mathbf{1}+\overline{\mathbf{5}}$.
However, the generators of the algebra of the first representation
are null, so there are just two embeddings of $SU(5)$ in $SU(6)$
: $\mathbf{6}\rightarrow\mathbf{1}+\mathbf{5}$ and $\mathbf{6}\rightarrow\mathbf{1}+\overline{\mathbf{5}}$.
Next, take the $SU(5)\rightarrow G_{SM}$ example. The fundamental
representation of $SU(5)$ must break into at least one non-trivial
representation of $SU(3)$, $SU(2)$, and the hypercharge group; otherwise
one or more of the generators of the algebra of $G_{SM}$ would be
null. So there are just two possibilities: $\mathbf{5}\rightarrow d^{c}+L$
or $\mathbf{5}\rightarrow\left(d^{c}\right)^{*}+L^{*}$;\footnote{Note that the hypercharges are fixed (up to some normalization factor
which we ignore here) by the tracelessness of the generators of the
algebra. This in turn is a consequence of the unitarity of the representation.} cases such as $\mathbf{5}\rightarrow\left(\mathbf{1},\mathbf{5},0\right)$
or $\mathbf{5}\rightarrow\left(\mathbf{1},\mathbf{2},y\right)+\left(\mathbf{1},\mathbf{2},y'\right)+\left(\mathbf{1},\mathbf{1},-2y-2y'\right)$
do not exist since they would require that one or more subgroup algebra
generators are the null matrices.

Based on these comments, all branching rules of $SU(N)$ into $SU(3)\times SU(2)\times U\left(1\right)^{m}$
can be found with the following algorithm:
\begin{enumerate}
\item Start by listing all irreducible representations of $SU(3)\times SU(2)$
up to size $N$.
\item Adding together the irreducible representations of $SU(3)\times SU(2)$
in all possible combinations, we get a complete list of the $N$-dimensional
representations of this group.
\item Exclude those $N$-dimensional representations of $SU(3)\times SU(2)$
which transform trivially either under $SU(3)$ or $SU(2)$. Furthermore,
discard the $N$-dimensional irreducible representations of $SU(3)\times SU(2)$
since this would force the charge operator of any additional $U(1)$
factor group to be null.
\item Finally, one must deal with the $U\left(1\right)^{m}$ charges of
these $N$-dimensional representations $R$ of $SU(3)\times SU(2)$,
which must be associated to traceless matrices. Let us assume that
the dimensions of the $n$ irreducible representations of $SU(3)\times SU(2)$
in a given $R$ and their charges under some $U(1)$ are $d\equiv\left(d_{1},d_{2},\cdots,d_{n}\right)$
and $q\equiv\left(q_{1},q_{2},\cdots,q_{n}\right)$ respectively.
Any $q$ vector is fine as long as $d\cdot q=0$, so immediately we
see that the maximal number of $U(1)$'s, which we have been calling
$m$ for some time, is exactly $n-1$ since this is the maximum number
of independent vectors orthogonal to $d$. Breaking $SU(3)\times SU(2)\times U\left(1\right)^{m}$
further down to the SM group requires, as previously explained, forming
the hypercharge group from a linear combination (any) of the $m=n-1$
$U(1)$'s obtained in this way.
\end{enumerate}
Among all these embeddings, there is one which is particularly interesting
(it exists for all $N\geq5$): under it, the anti-fundamental representation
$\overline{F}$ of $SU(N)$ breaks into the SM representations $d^{c}+L+(N-5)N^{c}$
(the number $N$ is not to be confused with $N^{c}$, which stands
for the $\left(\mathbf{1},\mathbf{1},0\right)$ representation of
$G_{SM}$). It is easy to check that the irreducible representation
of size $N\left(N-1\right)/2$ obtained from the anti-symmetric product
of two $F$'s (let us call it $K$ here) decomposes as
\begin{align}
K & \rightarrow Q+u^{c}+e^{c}+\left(N-5\right)\left(d^{c}\right)^{*}+\left(N-5\right)L^{*}+\frac{\left(N-5\right)\left(N-6\right)}{2}N^{c}\,,\label{eq:GSMemb1}
\end{align}
so the combination $-3\left(N-4\right)F+3K$ has the correct chirality.
In fact, for $N$ between 6 and 12, it turns out that this embedding
is the only one which can reproduce the SM chirality; this is the
conclusion of a scan over all combinations of fermion representations
up to some limiting size (at least 1000), considering as well in
the process all possible ways of forming the hypercharge operator
from the available $U(1)$'s. In \hyperref[sec:Appendix-A]{appendix}
there are various tables with the fermion content allowed by the chirality
constraint in these $SU(N)$ GUTs (the combination $-3\left(N-4\right)F+3K$
is not unique).\\

As the size of the $SU(N)$ groups increase, it becomes harder to
manage the problem of testing all the possible ways of forming the
SM hypercharge group. Yet, we should recall that for all groups $G_{GUT}$
where such an analysis was carried out --- $SO(10)$, $E_{6}$ or
$SU(12\geq n\geq5)$ --- it turned out that there was at most one
branching rule that worked for each $G_{GUT}$, and it matched the
one obtained by considering the $G_{SM}$ inside one of the $SU(5)$
subgroups of $G_{GUT}$. So, with hindsight, we could have simply
looked for the possible ways of embedding $SU(5)$ in the unification
groups, and then break $SU(5)$ to $G_{SM}$; no important embedding
of $G_{SM}$ in $G_{GUT}$ would have been missed. Or, better yet,
since an $SU(5)$ theory will only reproduce the SM's chirality with
the field combination $-3\left(\mathbf{5}\right)+3\left(\mathbf{10}\right)$,
we would just need to see if this $SU(5)$ chirality\footnote{By `$SU(5)$ chirality' we are referring to the concept of chirality
as discussed in section \eqref{sec:Framework}, but based on the analysis
of $SU(5)$ representations instead of those of $SU(3)\times SU(2)\times U(1)$.} is attainable with a given unification group.

Inspired by this observation, we have analyzed the different ways
of embedding $SU(5)$ in $SU(N\geq13)$. From an $SU(5)$ perspective,
the embedding in equation \eqref{eq:GSMemb1} corresponds to the following
branching rules,
\begin{align}
F & \rightarrow\mathbf{5}+\left(N-5\right)\mathbf{1}\,,\label{eq:SU5emb1_eq1-1}\\
K & \rightarrow\mathbf{10}+\left(N-5\right)\mathbf{5}+\frac{\left(N-5\right)\left(N-6\right)}{2}\mathbf{1}\,,\label{eq:SU5emb1_eq2-1}
\end{align}
where $F$ is the fundamental representation of $SU(N)$, and $K$
is the one in the anti-symmetric part of the product $F\times F$
as before. The field combination$-3\left(N-4\right)F+3K$ works from
the chirality point of view, and one might add that it does not lead
to gauge anomalies \cite{Adler:1969gk,Bell:1969ts}: indeed, for this
embedding, the chirality condition implies that the $SU(N)$ gauge
anomalies cancel, so all the configurations one can build from the
tables in \hyperref[sec:Appendix-A]{appendix} are fine.

Going through the $SU(N)$ family of groups, we have checked that
the embedding in equations \eqref{eq:SU5emb1_eq1-1} and \eqref{eq:SU5emb1_eq2-1}
is the only one that works. Until $SU(15)$ is reached. For this and
bigger groups, two new remarkable $SU(5)$ embeddings become possible.
The first one corresponds to
\begin{align}
F & \rightarrow2\left(\mathbf{5}\right)+\overline{\mathbf{5}}+\left(N-15\right)\mathbf{1}\,,
\end{align}
which means that $K\equiv\left(F\times F\right)_{\textrm{Ant.}}$
and $L\equiv\left(F\times F\right)_{\textrm{Sym.}}$ will break as
follows:
\begin{align}
K & \rightarrow2\left(\mathbf{24}\right)+\mathbf{15}+3\left(\mathbf{10}\right)+\overline{\mathbf{10}}+2\left(N-15\right)\mathbf{5}+\left(N-15\right)\overline{\mathbf{5}}+\left(\frac{244-31N+N^{2}}{2}\right)\mathbf{1}\,,\\
L & \rightarrow2\left(\mathbf{24}\right)+3\left(\mathbf{15}\right)+\overline{\mathbf{15}}+\mathbf{10}+2\left(N-15\right)\mathbf{5}+\left(N-15\right)\overline{\mathbf{5}}+\left(\frac{214-29N+N^{2}}{2}\right)\mathbf{1}\,.
\end{align}
As such, the field combination $-\left(N-12\right)F+2K-L$ will have
the correct chirality and it is anomaly free. This is not the unique
configuration that works for a given $N\geq15$; there are non-trivial
combinations of the representations $SU(N)$ which have no chirality,
just as in the embedding in equations \eqref{eq:SU5emb1_eq1-1}--\eqref{eq:SU5emb1_eq2-1}.
Nevertheless, we shall not print them here (they are fairly elaborate).

The second new noteworthy embedding encountered for $N\geq15$ is
the one under which the anti-fundamental representation of $SU(N)$
breaking into exactly one SM family plus singlets:
\begin{align}
\overline{F} & \rightarrow\overline{\mathbf{5}}+\mathbf{10}+\left(N-15\right)\mathbf{1}\,.\label{eq:emb3}
\end{align}
Obviously, the field configuration $-3F$ will have the correct chirality.
However, the reader will immediately notice that it leads to a theory
with an $SU(N)^{3}$ anomaly, which should be seen as a warning that
the chirality condition does not always imply absence of anomalies
(see \cite{Cahn:1981ub}). For this embedding, there are non-trivial
combinations $n_{i}$ of the $SU(N)$ representations with no chirality
which contribute to the $SU(N)^{3}$ anomaly, so one could have hoped
that for some $c_{i}$, the combination $-3F+\sum_{i}c_{i}n_{i}$
would be anomaly free. Unfortunately, no such $c_{i}$ exist for the
$SU(N)$ groups tested. How does one judge such models then? They
can be seen as non-renormalizable effective models \cite{Preskill:1990fr};
or maybe there is some way to cancel the anomaly (perhaps string inspired
\cite{Green:1984sg}). In any case, we shall not worry about these
anomalies --- we we simply assume that these models (or variations
of them) can be made part of a consistent quantum field theory. In
this spirit, we shall say a few more words about the curious embedding
in equation \eqref{eq:emb3}. Under it,
\begin{align}
K & \rightarrow\mathbf{45}+\overline{\mathbf{45}}+\left(N-15\right)\mathbf{5}+\overline{\mathbf{5}}+\left(N-15\right)\overline{\mathbf{10}}+\mathbf{10}+\frac{\left(N-15\right)\left(N-16\right)}{2}\mathbf{1}\,,
\end{align}
which means that the $\overline{K}$ representation of $SU(16+N')$
contains precisely $N'$ SM families plus vector particles. In the
particular case of $SU(19)$, looking through its $SO(10)$ subgroup
makes things even more clear:
\begin{align}
\overline{F} & \rightarrow\mathbf{16}+3\left(\mathbf{1}\right)\,,\\
\overline{K} & \rightarrow3\left(\mathbf{16}\right)+\mathbf{120}+3\left(\mathbf{1}\right)\,.
\end{align}
In other words, it is possible to unify the three SM families in $\overline{K}$
of $SU(19)$ (and more generally $N'$ families in $SU(16+N')$) although
the model will have gauge anomalies.

\begin{figure}[h]
\begin{centering}
\includegraphics[scale=0.35]{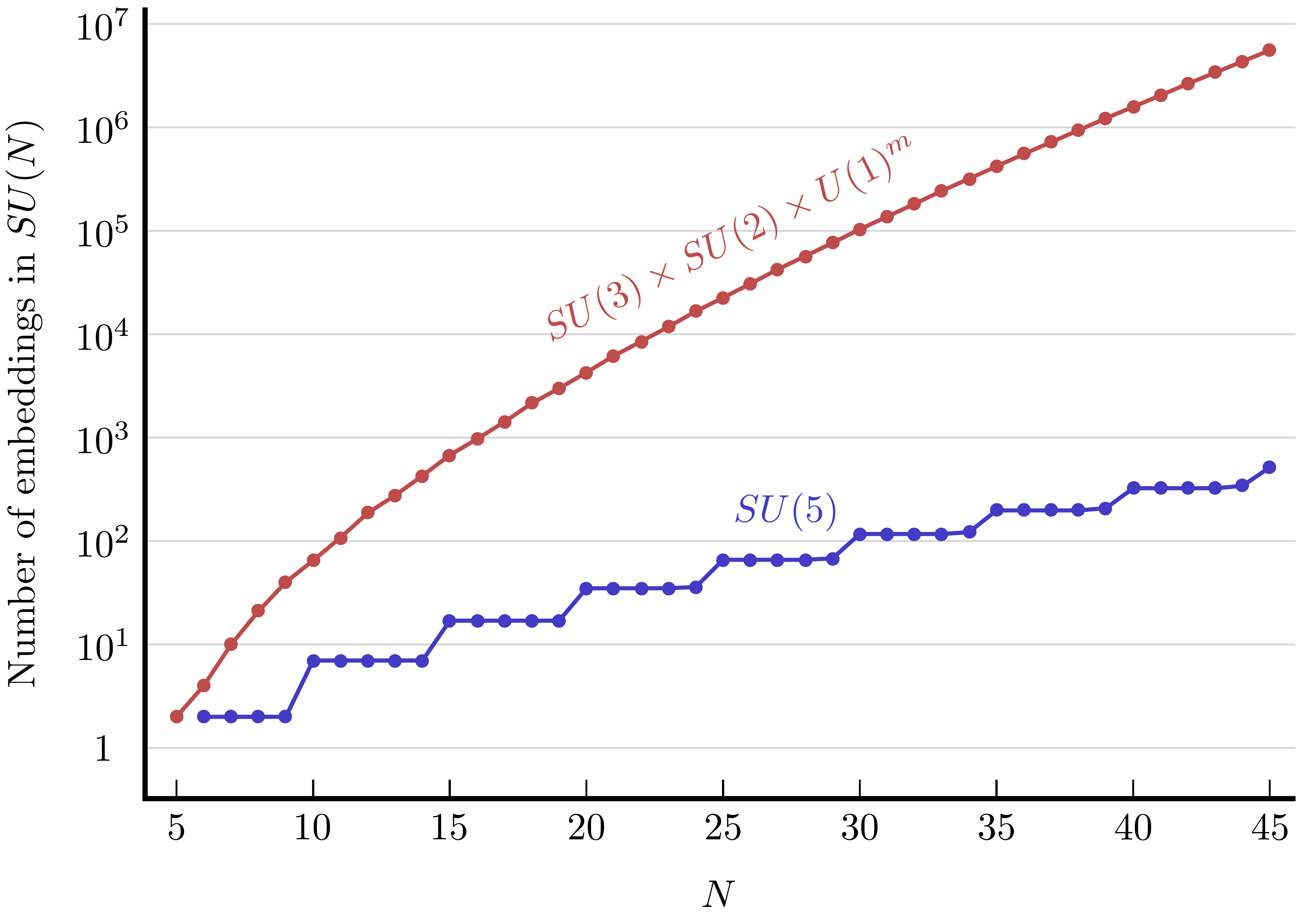}~~~~~~~~~~
\par\end{centering}

\protect\caption{\label{fig:Embedding}Number of distinct ways of embedding $SU(3)\times SU(2)\times U\left(1\right)^{m\geq1}$
and $SU(5)$ in $SU(N)$ groups. The counting includes only those
cases where $SU(3)\times SU(2)\times U\left(1\right)^{m}$ is embedded
in such a way that it is not a subgroup of a bigger $SU(3)\times SU(2)\times U\left(1\right)^{m+1}$
also contained in $SU(N)$. In this sense, $m$ must be maximal, yet
this does not mean that its value is fixed for a given $N$, as explained
in the main text.}
\end{figure}

The three embeddings of $SU(5)$ in $SU(N)$ which were presented
above are the only ones that work for $N\leq20$ (representations
of size up to one million were considered).  For $N>20$ one will
certainly find new embeddings: for example, starting with the $SU(45)$
group it is certainly possible to fit the three families (plus vector
particles) in the fundamental representation (although leading to
gauge anomalies once again). Yet, with increasingly big unification
groups the possibilities of embedding $SU(3)\times SU(2)\times U\left(1\right)^{m}$
in it grow in a seemingly exponential way (see figure \ref{fig:Embedding}).
For this reason, one might argue that models based on very large gauge
groups are not as attractive as those based on smaller ones: they
contain many subgroups, therefore a significant tuning of the scalar
sector parameters would likely be needed in order to have the correct
symmetry breaking.

~

\subsection{$SO(10+4N)$ with $N>0$}

In four dimensions and without confining interactions, is it possible
to embed the SM in a theory based on a gauge group of the family $SO(10+4N)$
for $N>0$? These groups do have complex representations and, furthermore,
it is possible to chirally embed the SM group in them (although such
embeddings are not plentiful --- see table \ref{tab:NumberOfGSMEmbbedings}).
However, despite these promising features, after performing computer
scans, it seems impossible to reproduce the SM chirality in $SO(14)$
and $SO(18)$ --- at least not with fermions representations with
a size smaller than 2 million.  Perhaps it has to do with the fact
that, unlike $SO(10)$, it is not possible to chirally embed $SU(5)$
in $SO(14)$ nor $SO(18)$. Interestingly, this last statement no
longer holds for bigger groups of the $SO(10+4N)$ family.

As with the special unitary groups, it becomes hard to analyze all
possible ways of embedding $G_{SM}$ so we shall focus instead on
the $SU(5)$ subgroups of $SO(10+4N)$ in the following.\footnote{The algorithm mentioned previously to find quick and easily the embeddings
of a subgroup $H$ in the special unitary groups cannot be readily
adapted to the special orthogonal groups. The method consisted essentially
in building all the $N$-dimensional representations of some $H\subset SU(N)$.
With $SO(N)$, one would have to consider only the strictly real $N$-dimensional
representations $R$ of $H\subset SO(N)$ (i.e., exclude the pseudo-real
and complex ones): for every such $R$ there in an embedding under
which the fundamental representation $F$ of $SO(N)$ breaks into
$R$.

But the branching rules of the spinor representation $S$ of $SO(N)$
would be missing. One can only speculate that perhaps one can find
these, up to conjugation, from the branching rules of the fundamental
representation $F$, since $F$ and $S$ are related. For example,
using the shorthand notation $X^{\left\{ m\right\} }$ and $X^{\left[m\right]}$
to denote the completely symmetry and anti-symmetric parts of the
product of $m$ copies of a representation $X$, there is the relation
$F^{\left[N\right]}+F^{\left[N-4\right]}+F^{\left[N-8\right]}+\cdots=S^{\left\{ 2\right\} }+\overline{S}^{\left\{ 2\right\} }$.} Table \ref{tab:SU5_embedding} contains a curious piece of information:
while there are two ways of embedding $SU(5)$ in $SO(10)$ --- a
pair of chiral embeddings --- there is just one for $SO(14)$ and
$SO(18)$, and under it the complex representations of these groups
break into a mixture of real and pairs of complex conjugated representations
of $SU(5)$. The number of embeddings is bigger for $SO(22)$ and
$SO(26)$ but they too are all vector embeddings. However, in this
case persistence pays off as it is possible to chirally embed $SU(5)$
in $SO(30)$.
\begin{table}
\begin{centering}
\begin{tabular}{ccccc}
\toprule 
Groups & \# Embeddings &  & Groups & \# Embeddings\tabularnewline
\cmidrule{1-2} \cmidrule{4-5} 
$SO(10)$ & 2 (1) &  & $SO(26)$ & 4 (0)\tabularnewline
$SO(14)$, $SO(18)$ & 1 (0) &  & $SO(30)$ & 10 (3)\tabularnewline
$SO(22)$ & 3 (0) &  &  & \tabularnewline
\bottomrule
\end{tabular}
\par\end{centering}

\protect\caption{\label{tab:SU5_embedding}Number of district ways of embedding $SU(5)$
in the groups $SO(10+4N)$ with $0\leq N\leq5$. Only some of these correspond to chiral embeddings (the integers in parenthesis indicate the number of pairs of chiral embeddings in each case).}
\end{table}

The trouble with the $SO(10+4N)$ group family is that the complex
representations become exponentially large with $N$: the spinor representation
--- which is the smallest one --- has $4^{2+N}$ components. In the
case of $SO(30)$, we have considered the complex representations
smaller than the $\mathbf{132562944}$ (there are only 7) and all
the chiral embeddings of $SU(5)$ in $SO(30)$; it turns out that
it is impossible to obtain three families of $\overline{\mathbf{5}}+\mathbf{10}$
plus vector particles.

As for bigger groups in the $SO(10+4N)$ family, they were not tested
so one can only speculate about the possibility of embedding the three
SM families in such models. The complex representations will be even
bigger than those of $SO(30)$, each of them potentially breaking
into thousands of distinct complex $SU(5)$ representations, so it
seems unlikely that one could match all these sub-representations
in pairs $\left(R,\overline{R}\right)$ leaving only a small excess
of $\overline{\mathbf{5}}$'s and $\mathbf{10}$'s over $\mathbf{5}$'s
and $\overline{\mathbf{10}}$'s. Even if it is possible, it would
almost inevitably require millions of new vector particle components.\footnote{To have a feeling of the huge numbers involved, the three smallest
complex representations of $SO(30)$ have sizes 16384, 475136, 6635520;
in $SO(34)$ these numbers increase fourfold or more to 65536, 2162688,
34537472.}

\subsection{$SU(3)\times SU\left(3\right)$ and $SU(3)\times SU\left(3\right)\times U(1)$}

We shall consider $SU(3)\times SU(3)$ --- even though it is not a
simple group --- because it contains $G_{SM}$ as a subgroup, it has
a minimal number of diagonal generators, and it is a group with complex
representations. Besides $SU(5)$, the only other semi-simple group
with these properties is $SU(3)\times SU(2)\times SU(2)$, which clearly
will not yield the correct chirality as one would always have pairs
of representations with opposite hypercharges. Are there $SU(3)\times SU(3)$
models with the correct chirality? Models with an extra $U(1)$ can
successfully embed the SM, achieving the correct chirality, but without
this extra abelian factor we shall see that this is not possible.

One of the two factors is the color group $SU(3)_{C}$ which must
not be broken, while the other --- let us call it henceforth $SU(3)_{L}$
--- has to break into the EW group $SU(2)_{L}\times U(1)_{Y}$. We
recall here that the representations of $SU(3)_{L}$ can be uniquely
labeled by two non-negative integers $\left\{ a,b\right\} $ (the
Dynkin coefficients) which in terms of Young tableaux can be identified
with the representation with $a$ columns with a single row and $b$
columns with two rows. The conjugate representation of $\left\{ a,b\right\} $
is $\left\{ b,a\right\} ,$ and their dimension is $\frac{1}{2}\left(a+1\right)\left(b+1\right)\left(a+b+2\right)$
--- for example, $\left\{ 0,0\right\} $, $\left\{ 1,0\right\} $
and $\left\{ 0,1\right\} $ are the singlet, triplet (by convention),
and anti-triplet representations, respectively. On the other hand,
the $SU(2)_{L}\times U(1)_{Y}$ representations can be labeled by
their spin $j$ and (unnormalized) hypercharge $y$.

Crucially, it is possible to derive the branching rules of a generic
representation of $SU(3)_{L}$ into those of $SU(2)_{L}\times U(1)_{Y}$:
$\left\{ a,b\right\} $ decomposes into representations with hypercharge
$y=a-b+3n$ ($n\in\mathbb{Z},-a\leq n\leq b$) whose $SU(2)_{L}$
spins are $j=\frac{a+b-n}{2},\frac{a+b-n}{2}-1,\frac{a+b-n}{2}-2,\cdots,\left|\frac{b-a-n}{2}\right|$
(for $n\geq0$) and $j=\frac{a+b+n}{2},\frac{a+b+n}{2}-1,\frac{a+b+n}{2}-2,\cdots,\left|\frac{b-a+n}{2}\right|$
(for $n<0$).\footnote{For example, $\left\{ 3,2\right\} $ decomposes into $\left(j,y\right)=$
$\left(\frac{3}{2},7\right)$, $\left(1\oplus2,4\right)$, $\left(\frac{1}{2}\oplus\frac{3}{2}\oplus\frac{5}{2},1\right)$,
$\left(0\oplus1\oplus2,-2\right)$, $\left(\frac{1}{2}\oplus\frac{3}{2},-5\right)$,
and $\left(1,-8\right)$.} The state with the biggest spin, $\left(j,y\right)=\left(\frac{a+b}{2},a-b\right)$,
is a complex $SU(2)_{L}\times U(1)_{Y}$ representation whenever the
$SU(3)_{L}$ representation from which it originates is complex as
well ($a\neq b$). On the other hand, if $a+b>1$ this state will
be an $SU(2)$ triplet or higher-dimensional representation which
must have a vector mass (since it is not seen at low energies). As
such, in a model where there is an $\left\{ a,b\right\} $ complex
representation of $SU(3)_{L}$ with $a+b>1$, one needs to ensure
the presence of at least another $SU(3)_{L}$ representation which
contains the state $\left(j,y\right)=\left(\frac{a+b}{2},b-a\right)$.
Obviously this can be achieved with the representation $\left\{ a,b\right\} ^{*}=\left\{ b,a\right\} $,
but having both $\left\{ a,b\right\} $ and its conjugate would not
affect a model's chirality (see section \ref{sec:Framework}) and
for that reason we have stated previously that such configurations
should be excluded from the analysis. Therefore, in order to obtain
the $SU(2)_{L}\times U(1)_{Y}$ representation $\left(j,y\right)=\left(\frac{a+b}{2},b-a\right)$
one must have an $SU(3)_{L}$ complex representation $\left\{ a',b'\right\} $
with $a'+b'>a+b$. But such a $\left\{ a',b'\right\} $ multiplet
would decompose into complex states with spin even higher than $\frac{a+b}{2}$,
presenting a renewed problem. And thus, with this circular argument
it is shown that by using any $SU(3)_{L}$ complex fermion representation
other that $\left\{ 0,0\right\} $ (the singlet), $\left\{ 1,0\right\} $
(the triplet) and $\left\{ 0,1\right\} $ (the anti-triplet) it will
be impossible to avoid having fermions in triplet or higher dimensional
representations of $SU(2)_{L}$ with no vector mass. And yet, with
just $SU(3)_{L}$ singlets, triplets and anti-triplet it is not possible
to reproduce the EW representations of the SM fermions, therefore
one cannot embed the SM in an $SU(3)_{C}\text{\ensuremath{\times}}SU(3)_{L}$
based model.

~

Viable models can be built by adding an extra $U(1)_{X}$ factor,
where the SM hypercharge $Y$ is a combination of the $T_{8}$ generator
of $SU(3)_{L}$ and $X$:
\begin{align}
Y & \propto\frac{\kappa}{\sqrt{3}}T_{8}+X\,,
\end{align}
for some $\kappa$ factor which controls the relative weight of $X$
in $Y$. A known possibility is to take $\kappa=1$, with the leptons
in the representations $3\left(\boldsymbol{1},\overline{\boldsymbol{3}},-\frac{1}{3}\right)+3\left(\boldsymbol{1},\boldsymbol{1},1\right)$
and the quarks in $2\left(\boldsymbol{3},\boldsymbol{3},0\right)+\left(\boldsymbol{3},\overline{\boldsymbol{3}},\frac{1}{3}\right)+4\left(\overline{\boldsymbol{3}},\boldsymbol{1},-\frac{2}{3}\right)+5\left(\overline{\boldsymbol{3}},\boldsymbol{1},\frac{1}{3}\right)$
\cite{Singer:1980sw}; an alternative is to take $\kappa=-3$, placing
the leptons in the representations $3\left(\boldsymbol{1},\boldsymbol{3},0\right)$
and the quarks in $2\left(\boldsymbol{3},\overline{\boldsymbol{3}},-\frac{1}{3}\right)+\left(\boldsymbol{3},\boldsymbol{3},\frac{2}{3}\right)+3\left(\overline{\boldsymbol{3}},\boldsymbol{1},-\frac{2}{3}\right)+3\left(\overline{\boldsymbol{3}},\boldsymbol{1},\frac{1}{3}\right)+2\left(\overline{\boldsymbol{3}},\boldsymbol{1},\frac{4}{3}\right)+\left(\overline{\boldsymbol{3}},\boldsymbol{1},-\frac{5}{3}\right)$
\cite{Pisano:1991ee,Frampton:1992wt,Foot:1992rh}. Trivial variations
to the fermion content --- adding real or conjugate pairs of representations
of the $SU(3)_{C}\times SU(3)_{L}\times U(1)_{X}$ gauge group ---
are always possible and might even have interesting motivations (see
for example \cite{Boucenna:2014dia}). But what about non-trivial
variations? Just as with simple unification groups, the need to reproduce
the SM chirality poses a very significant constraint. We do note that
because of the extra $U(1)_{X}$ group factor the chirality condition
does not imply automatically the absence of gauge anomalies, so these
should be seen as complementary conditions on the possible fermion
content of a given model.

We shall state here what are the simplest\footnote{The criteria for simplicity adopted here is based on the size of the
$SU(3)_{L}$ representations involved: solutions with smaller representations
are taken to be simpler.} non-trivial modifications which can be made to the known $SU(3)_{C}\times SU(3)_{L}\times U(1)_{X}$
models. If $R$ is a complex representation of $SU(3)_{C}$, then
it turns out that the simplest combinations of fields with no chirality
and no gauge anomalies is of the form\footnote{We stress again the content of footnote \ref{fn:Notation_comment}
concerning the naming of representations. This implies in particular
that the $\overline{\boldsymbol{6}}$ of $SU(3)$ in this work is
the $\boldsymbol{6}$ used in reference \cite{Slansky:1981yr}, and
vice-versa.}
\begin{align}
 & -4\left(R,\boldsymbol{1},z\right)+5\left(R,\boldsymbol{3},\frac{4}{3}\kappa+z\right)+5\left(R,\overline{\boldsymbol{6}},\frac{2}{3}\kappa+z\right)-5\left(R,\boldsymbol{8},\kappa+z\right)\nonumber \\
 & \quad\quad\quad\quad\quad\quad\quad\quad\quad\quad\quad\quad\quad-\left(R,\boldsymbol{10},2\kappa+z\right)-\left(R,\boldsymbol{15'},\frac{4}{3}\kappa+z\right)+\left(R,\overline{\boldsymbol{24}},\frac{5}{3}\kappa+z\right)\,,\label{eq:331_Rcomplex}
\end{align}
for some arbitrary number $z$ ($\kappa$ was introduced earlier).
Remarkably, this expression introduces many new $SU(3)_{L}$ representations
going all the way up to size 24. So, for example, in the model $SU(3)_{C}\times SU(3)_{L}\times U(1)_{X}$
with $\kappa=1$ \cite{Singer:1980sw}, if we were to replace the
quarks in four copies of the representation $\left(\overline{\boldsymbol{3}},\boldsymbol{1},-\frac{2}{3}\right)$
by something else, the simplest possibility would be (using equation
\eqref{eq:331_Rcomplex} with $R=\overline{\boldsymbol{3}}$, $z=-\frac{2}{3}$)
\begin{equation}
4\left(\overline{\boldsymbol{3}},\boldsymbol{1},-\frac{2}{3}\right)\mapsto5\left(\overline{\boldsymbol{3}},\boldsymbol{3},\frac{2}{3}\right)+5\left(\overline{\boldsymbol{3}},\overline{\boldsymbol{6}},0\right)+5\left(\boldsymbol{3},\boldsymbol{8},-\frac{1}{3}\right)+\left(\boldsymbol{3},\overline{\boldsymbol{10}},-\frac{4}{3}\right)+\left(\boldsymbol{3},\overline{\boldsymbol{15'}},-\frac{2}{3}\right)+\left(\overline{\boldsymbol{3}},\overline{\boldsymbol{24}},1\right)\,.
\end{equation}
The complexity of this mixture of representations can be interpreted
as saying that the quark assignment in $SU(3)_{C}\times SU(3)_{L}\times U(1)_{X}$
models (for a fixed $\kappa$) is, for practical purposes, unique.
Note that gauge anomaly cancellation alone would allow the replacement
of $4\left(\overline{\boldsymbol{3}},\boldsymbol{1},-\frac{2}{3}\right)$
with much simpler combinations.

If $R$ is a real representation of $SU(3)_{C}$, the combination
in equation \eqref{eq:331_Rcomplex} is still valid, but it is not
the simplest one. That distinction goes to
\begin{align}
 & \left(R,\boldsymbol{3},x\right)+\left(R,\boldsymbol{3},-\frac{\kappa}{3}-x\right)-\left(R,\boldsymbol{3},y\right)-\left(R,\boldsymbol{3},-\frac{\kappa}{3}-y\right)+\left(R,\boldsymbol{1},\frac{\kappa}{3}-x\right)+\nonumber \\
 & \quad\quad\quad\quad\quad\quad\quad\quad\quad\quad\quad\quad\quad\quad+\left(R,\boldsymbol{1},\frac{2}{3}\kappa+x\right)-\left(R,\boldsymbol{1},\frac{\kappa}{3}-y\right)-\left(R,\boldsymbol{1},\frac{2}{3}\kappa+y\right)\,,
\end{align}
where $x$ should be different from $y$ and $-\frac{\kappa}{3}-y$,
otherwise this would be a self-conjugate combination of fields. Note
also that we can drop one of the last four representations if it is
made real by choosing an appropriate value for $x$ or $y$. As such,
referring once more to the model of \cite{Singer:1980sw} where $\kappa=1$,
one could replace the leptons in the three copies of the representation
$\left(\boldsymbol{1},\overline{\boldsymbol{3}},-\frac{1}{3}\right)$
by three copies of the following rather more complex combination,
which nevertheless does not involve $SU(3)_{L}$ representations bigger
than (anti-)triplets:
\begin{equation}
\left(\boldsymbol{1},\boldsymbol{3},-\frac{2}{3}\right)+\left(\boldsymbol{1},\overline{\boldsymbol{3}},-y\right)+\left(\boldsymbol{1},\overline{\boldsymbol{3}},\frac{1}{3}+y\right)+\left(\boldsymbol{1},\boldsymbol{1},1\right)+\left(\boldsymbol{1},\boldsymbol{1},-\frac{1}{3}+y\right)+\left(\boldsymbol{1},\boldsymbol{1},-\frac{2}{3}-y\right)\,.
\end{equation}

In summary then, for a fixed embedding of the SM group in $SU(3)_{C}\times SU(3)_{L}\times U(1)_{X}$,
it is fairly complicated to assign the leptons to other representations,
and even more so for the quarks.

\section{\label{sec:Discussion-and-concluding}Concluding remarks}

Pairs of right- and left-handed fermions transforming in the same
way under the SM gauge group (vector fermions) can have a very high
mass and therefore escape direct observation. On the other hand, unpaired
ones (chiral fermions) may only get a small mass after electroweak
symmetry breaking. Therefore, in Grand Unified Theories one must avoid
introducing chiral fermions beyond those present in the Standard Model.

Motivated by this observation, we have analyzed in a systematic way
the fermion sector of GUTs containing the SM fermions plus vector
particles only. This very simple requirement on the fermion content
of GUTs turns out to be a very constraining one (implying in most
cases, but not all, the cancellation of all gauge anomalies). The
analysis carried out assumes that there are no extra dimensions nor
confining gauge interactions at high energies.

A thorough computer scan was performed over all simple groups with
rank smaller than 12 (excluding $SO(22)$) and over all their representations
up to some size (from a few thousands up to millions, depending on
the group). A very significant part of the work consisted in tracking
down and looking at all possible ways of embedding $G_{SM}$ in each
unification group: this required cataloging the different ways that
$SU(3)\times SU(2)\times U(1)^{m}$ can be embedded in $G_{GUT}$
and, for each of them, to consider that the SM's $U(1)_{Y}$ can be
a priori any linear combination of the $m$ available $U(1)$'s.

With these simple groups, how exotic can GUTs be (the group, the embedding,
and the field content)? Concerning the group, $SO(14)$ and $SO(18)$
were found not to work, leaving $SO(10)$, $SU(5\leq N\leq12)$ and
$E_{6}$ as viable unification groups. Surprisingly, it was found
that the SM gauge group must be embedded in each of these groups such
that the GUT representations decompose in a unique way into SM fields.
This uniqueness is far from obvious, even though model builders have
been working with these groups and these embeddings for a long time.
In every case, it turns out that the SM group is embedded in the GUT
group in such a way that it can be viewed as going through $SU(5)$ 
for the calculation of the representation branching rules,
$G_{GUT}\rightarrow SU(5)\rightarrow G_{SM}$, so an important consequence
of this result is that the hypercharge normalization factor $\sqrt{3/5}$,
usually associated to $SU(5)$ GUTs, is in fact universal.\footnote{For the extensive list of groups and representations considered at
least.} Indeed, the relations $g_{1}=\sqrt{5/3}g'$ , $g_{2}=g$ and $g_{3}=g_{s}$
are the correct ones for all the tested cases.

As far as the field content is concerned, we dismissed the introduction
of real fermion GUT representations or pairs of complex conjugate
ones as trivial variations to the fermion sector of a model. It is
also inconsequential to exchange the GUT representations $\sum_{i}R_{i}$
by their conjugates $\sum_{i}\overline{R_{i}}$, since one will recover
the same SM representations by considering a different embedding of
$G_{SM}$ in $G_{GUT}$. Factoring out such variations, the standard
fermion assignments $3\left(\overline{\mathbf{5}}\right)+3\left(\mathbf{10}\right)$
in $SU(5)$ and $3\left(\mathbf{27}\right)$ in $E_{6}$ appear to
be unique, while $3\left(\mathbf{16}\right)$ in $SO(10)$ is not.
In this last case, it is possible to get rid of the spinor representation,
but that requires the introduction of complicated mixtures of very
large representations. On the other hand, $SU(5<N\leq12)$ models
may have a rich variety of different fermion representations, some
of which have been already explored in the literature. For example,
it is possible to have an $SU(11)$ model with just three fermion
representations, $2\left(\overline{\mathbf{55}}\right)+\mathbf{462}$,
matching in this sense the minimality of $SO(10)$ and $E_{6}$ GUTs,
and perhaps exhibiting an interesting flavor structure.

Two non-simple groups were looked at as well. It was shown that no
model based on the gauge group $SU(3)\times SU\left(3\right)$ can
yield the SM chirality. On the other hand, viable models are known
to exist with an extra $U(1)$, and in this work we commented that
non-trivial changes to their fermion sector are possible, although
they do need to be very elaborate.

Bigger unification groups were also considered, assuming that $G_{SM}$
is in an $SU(5)$ subgroup of $G_{GUT}$. This analysis strongly suggests
that $SO(10<N\leq30)$ are not suitable grand unified groups, in contrast
to the $SU(N)$'s which, for $N\geq15$, can actually embed $G_{SM}$
in multiple valid ways. Interestingly, under one of these embeddings
it is possible to unify $N'$ SM families in a single representation
of $SU(16+N')$. In particular, the $\mathbf{171}$ of $SU(19)$ contains
the SM fermions plus vector particles only. However, one should keep
in mind that family unification with special unitary groups leads
to gauge anomalies which, in a fundamental theory, need to be dealt
with.

\subsection*{\vspace{-0.15cm}Acknowledgments}

The author wishes to thank Sofiane Boucenna, Martin Hirsch, Michal Malinsk\'y, and Jos\'e
Valle for discussions related to this document, as well as for comments
provided on a draft version of it. This work was supported by the
Spanish grants FPA2014-58183-P and Multidark CSD2009-00064 (from the
\textit{Ministerio de Econom\'ia y Competitividad}), PROMETEOII/2014/084
(from the \textit{Generalitat Valenciana}), as well as the Portuguese
grant EXPL/FIS-NUC/0460/2013 (from the \textit{Funda\c{c}\~ao para a Ci\^encia
e a Tecnologia}).

\clearpage

\appendix

\section*{\label{sec:Appendix-A}Appendix\addcontentsline{toc}{section}{\protect\numberline{}Appendix}}

As mentioned in the text, the embedding of the SM group in $SU(5<N\leq12)$
characterized by the branching rule in equation \eqref{eq:SU5emb1_eq1-1}
was the only one found to be capable of reproducing the SM chirality.
The fermion content $\widetilde{c}\equiv-3\left(N-4\right)F+3K$ is
a particular example of a valid one ($F$ being the fundamental representation
of $SU(N)$ and $K\equiv\left(F\times F\right)_{\textrm{Ant.}}$).
Other equality valid combinations of fields can be built by adding
non-trivial combinations $n_{i}$ of the $SU(N)$ representations
with no chirality (as discussed in section \eqref{sec:Framework}).
This appendix contains the tables \eqref{tab:SU6}--\eqref{tab:SU12}
which list all such independent $n_{i}$, for $5<N\leq12$. The biggest
representation considered for the elaboration of each table was determined
mainly by space considerations, given that more solutions seem to
always appear if one includes bigger ones. For each group, one or
more ``interesting solutions'' were picked in a somewhat subjective
manner: they are notable for having either a reduced number of types
of representations or a small number of total representations (including
multiplicity).

\begin{center}
\begin{table}[h]
\begin{centering}
\begin{tabular}{cc}
\toprule 
\begin{tabular}{c}
Maximum size of\tabularnewline
representations considered\tabularnewline
\end{tabular} & 1000\tabularnewline
 & \tabularnewline
\begin{tabular}{c}
Non-trivial combinations\tabularnewline
of representations with\tabularnewline
no chirality ($n_{i}$)\tabularnewline
\end{tabular} & $\begin{array}{c}
-\mathbf{6}+\mathbf{21}-\mathbf{70}+\mathbf{84}-\mathbf{105}\,,\\
-\mathbf{6}-\mathbf{15}+\mathbf{21}+\mathbf{84}+\mathbf{105'}-\mathbf{210}\,,\\
-\mathbf{21}+\mathbf{56}+\mathbf{120}+\mathbf{210'}-\mathbf{280}-\mathbf{336}\,,\\
-\mathbf{6}+\mathbf{21}-\mathbf{70}+\mathbf{84}-\mathbf{120}-\mathbf{210'}+\mathbf{280}\\
\quad\quad\quad\quad\quad\quad\quad+\mathbf{420}-\mathbf{560}-\mathbf{840}+\mathbf{840'}\,,\\
-\mathbf{56}-\mathbf{126}+\mathbf{315}+\mathbf{504}-\mathbf{720}+\mathbf{840''}\,,\\
-\mathbf{6}+\mathbf{56}-\mathbf{70}+\mathbf{384}+\mathbf{420}-\mathbf{840}+\mathbf{896}\,.
\end{array}$\tabularnewline
 & \tabularnewline
\begin{tabular}{c}
Interesting solutions \tabularnewline
with the SM chirality\tabularnewline
\end{tabular} & $\begin{array}{rl}
\widetilde{c}= & -6\repb 6+3\repb{15}\end{array}$\tabularnewline
\bottomrule
\end{tabular}
\par\end{centering}

\protect\caption{\label{tab:SU6}Information on the combinations of $SU(6)$ representations
yielding the correct chirality. For example, the simplest solution,
$\widetilde{c}$, together with more vector particles is discussed
in \cite{Berezhiani:1995dt,Shafi:2001iu,CarcamoHernandez:2010im}. }
\end{table}

\par\end{center}

\begin{center}
\begin{table}[h]
\begin{centering}
\begin{tabular}{cc}
\toprule 
\begin{tabular}{c}
Maximum size of\tabularnewline
representations considered\tabularnewline
\end{tabular} & 1000\tabularnewline
 & \tabularnewline
\begin{tabular}{c}
Non-trivial combinations\tabularnewline
of representations with\tabularnewline
no chirality ($n_{i}$)\tabularnewline
\end{tabular} & $\begin{array}{c}
-\mathbf{7}+\mathbf{21}-\mathbf{35}\,,\\
3\repb 7-3\repb{28}+2\repb{112}-\rep{140}+\rep{210}\,,\\
4\repb 7-2\repb{28}+\rep{112}-2\repb{140}+\rep{224}\,,\\
5\repb 7+3\repb{21}-4\repb{28}-2\repb{140}-2\repb{196}+\rep{490}\,,\\
\rep 7+4\repb{21}-\rep{28}-2\repb{112}-\rep{140}-2\repb{196}+\rep{490'}\,,\\
4\repb 7+4\repb{21}-3\repb{28}-3\repb{140}-\rep{196}+\rep{588}\,,\\
3\repb{28}-3\repb{84}-2\repb{189}-2\repb{378}+\rep{540}+\rep{756}\,,\\
4\repb{28}-2\repb{84}-3\repb{189}-\rep{378}+2\repb{540}+\rep{840}\,.
\end{array}$\tabularnewline
 & \tabularnewline
\begin{tabular}{c}
Interesting solutions \tabularnewline
with the SM chirality\tabularnewline
\end{tabular} & $\begin{array}{rl}
\widetilde{c}= & -9\left(\mathbf{\mathbf{7}}\right)+3\left(\mathbf{21}\right)\,,\\
\widetilde{c}-3n_{1}= & -6\left(\mathbf{\mathbf{7}}\right)+3\left(\mathbf{35}\right)\,.
\end{array}$\tabularnewline
\bottomrule
\end{tabular}
\par\end{centering}

\protect\caption{Information on the combinations of $SU(7)$ representations yielding
the correct chirality. For example, \cite{Chkareuli:2000bm} uses
$\widetilde{c}$ and \cite{Frampton:1979cw} mentions both $\widetilde{c}$
and $\widetilde{c}-3n_{1}$.}
\end{table}

\par\end{center}

\begin{center}
\begin{table}[h]
\begin{centering}
\begin{tabular}{cc}
\toprule 
\begin{tabular}{c}
Maximum size of\tabularnewline
representations considered\tabularnewline
\end{tabular} & 1500\tabularnewline
 & \tabularnewline
\begin{tabular}{c}
Non-trivial combinations\tabularnewline
of representations with\tabularnewline
no chirality ($n_{i}$)\tabularnewline
\end{tabular} & $\begin{array}{c}
-3\repb 8+2\repb{28}-\rep{56}\,,\\
6\repb 8-6\repb{36}+3\repb{168}-\rep{216}-\rep{378}\,,\\
10\repb 8-3\repb{36}+\rep{168}-3\repb{216}+\rep{420}\,,\\
15\repb 8-8\repb{36}+3\repb{168}-3\repb{216}+\rep{504}\,,\\
15\repb 8+6\repb{28}-10\repb{36}-3\repb{216}+3\repb{336}+\rep{1008}\,,\\
2\repb 8+17\repb{28}-3\repb{36}-7\repb{168}-3\repb{216}+5\repb{336}+\rep{1176}\,,\\
10\repb 8+10\repb{28}-6\repb{36}-6\repb{216}+\rep{336}+\rep{1344}\,.
\end{array}$\tabularnewline
 & \tabularnewline
\begin{tabular}{c}
Interesting solutions \tabularnewline
with the SM chirality\tabularnewline
\end{tabular} & $\begin{array}{rl}
\widetilde{c}= & -12\left(\mathbf{8}\right)+3\left(\mathbf{28}\right)\,,\\
\widetilde{c}-4n_{1}= & -5\left(\mathbf{28}\right)+4\left(\mathbf{56}\right)\,.
\end{array}$\tabularnewline
\bottomrule
\end{tabular}
\par\end{centering}

\protect\caption{Information on the combinations of $SU(8)$ representations yielding
the correct chirality. For example, in \cite{Barr:2008pn} the author
considers the combination $\widetilde{c}-n_{1}$ which is readily
seen to be the one involving the least amount of fermion components.
Reference \cite{Chkareuli:1992kd} considers instead the combination
$\widetilde{c}-2n_{1}$.}
\end{table}

\par\end{center}

\begin{center}
\begin{table}[h]
\begin{centering}
\begin{tabular}{cc}
\toprule 
\begin{tabular}{c}
Maximum size of\tabularnewline
representations considered\tabularnewline
\end{tabular} & 1500\tabularnewline
 & \tabularnewline
\begin{tabular}{c}
Non-trivial combinations\tabularnewline
of representations with\tabularnewline
no chirality ($n_{i}$)\tabularnewline
\end{tabular} & $\begin{array}{c}
-6\repb 9+3\repb{36}-\rep{84}\,,\\
-5\repb 9+2\repb{36}-\rep{126}\,,\\
10\repb 9-10\repb{45}+4\repb{240}-\rep{315}-\rep{630}\,,\\
20\repb 9-4\repb{45}+\rep{240}-4\repb{315}+\rep{720}\,,\\
36\repb 9-20\repb{45}+6\repb{240}-4\repb{315}+\rep{1008}\,,\\
45\repb 9-15\repb{45}+4\repb{240}-6\repb{315}+\rep{1050}\,.
\end{array}$\tabularnewline
 & \tabularnewline
\begin{tabular}{c}
Interesting solutions \tabularnewline
with the SM chirality\tabularnewline
\end{tabular} & $\begin{array}{rl}
\widetilde{c}= & -15\left(\mathbf{9}\right)+3\left(\mathbf{36}\right)\,,\\
\widetilde{c}-n_{1}= & -9\left(\mathbf{9}\right)+\mathbf{84}\,,\\
\widetilde{c}-3n_{2}= & -3\left(\mathbf{36}\right)+3\left(\mathbf{126}\right)\,.
\end{array}$\tabularnewline
\bottomrule
\end{tabular}
\par\end{centering}

\protect\caption{Information on the combinations of $SU(9)$ representations yielding
the correct chirality. For example, the solutions $\widetilde{c}-n_{1}$
and $\widetilde{c}-n_{1}+n_{2}$ are mentioned in \cite{Frampton:1979tj}
and \cite{Dent:2009pd} respectively.}
\end{table}

\par\end{center}

\begin{center}
\begin{table}[h]
\begin{centering}
\begin{tabular}{cc}
\toprule 
\begin{tabular}{c}
Maximum size of\tabularnewline
representations considered\tabularnewline
\end{tabular} & 2000\tabularnewline
 & \tabularnewline
\begin{tabular}{c}
Non-trivial combinations\tabularnewline
of representations with\tabularnewline
no chirality ($n_{i}$)\tabularnewline
\end{tabular} & $\begin{array}{c}
-10\repb{10}+4\repb{45}-\rep{120}\,,\\
-16\repb{10}+5\repb{45}-\rep{210}\,,\\
15\repb{10}-15\repb{55}+5\repb{330}-\rep{440}-\rep{990}\,,\\
35\repb{10}-5\repb{55}+\rep{330}-5\repb{440}+\rep{1155}\,,\\
70\repb{10}-40\repb{55}+10\repb{330}-5\repb{440}-\rep{1848}\,,\\
105\repb{10}-24\repb{55}+5\repb{330}-10\repb{440}+\rep{1980}\,.
\end{array}$\tabularnewline
 & \tabularnewline
\begin{tabular}{c}
Interesting solutions \tabularnewline
with the SM chirality\tabularnewline
\end{tabular} & $\begin{array}{rl}
\widetilde{c}= & -18\left(\mathbf{10}\right)+3\left(\mathbf{45}\right)\,,\\
\widetilde{c}+3n_{1}-3n_{2}= & -3\left(\mathbf{120}\right)+3\left(\mathbf{210}\right)\,.
\end{array}$\tabularnewline
\bottomrule
\end{tabular}
\par\end{centering}

\protect\caption{Information on the combinations of $SU(10)$ representations yielding
the correct chirality. Reference \cite{Kashibayashi:1982nu} mentions
explicitly the solutions $\widetilde{c}$, $\widetilde{c}-n_{1}$
and $\widetilde{c}-2n_{1}$.}
\end{table}

\par\end{center}

\begin{center}
\begin{table}[h]
\begin{centering}
\begin{tabular}{cc}
\toprule 
\begin{tabular}{c}
Maximum size of\tabularnewline
representations considered\tabularnewline
\end{tabular} & 3000\tabularnewline
 & \tabularnewline
\begin{tabular}{c}
Non-trivial combinations\tabularnewline
of representations with\tabularnewline
no chirality ($n_{i}$)\tabularnewline
\end{tabular} & $\begin{array}{c}
-15\repb{11}+5\repb{55}-\rep{165}\,,\\
-35\repb{11}+9\repb{55}-\rep{330}\,,\\
-21\repb{11}+5\repb{55}-\rep{462}\,,\\
21\repb{11}-21\repb{66}+6\repb{440}-\rep{594}-\rep{1485}\,,\\
56\repb{11}-6\repb{66}+\rep{440}-6\repb{594}+\rep{1760}\,.
\end{array}$\tabularnewline
 & \tabularnewline
\begin{tabular}{c}
Interesting solutions \tabularnewline
with the SM chirality\tabularnewline
\end{tabular} & $\begin{array}{rl}
\widetilde{c}= & -21\left(\mathbf{11}\right)+3\left(\mathbf{55}\right)\,,\\
\widetilde{c}+n_{1}-n_{2}= & -\mathbf{11}-\mathbf{55}-\mathbf{165}+\mathbf{330}\,,\\
\widetilde{c}-n_{3}= & -2\left(\mathbf{55}\right)+\mathbf{462}\,.
\end{array}$\tabularnewline
\bottomrule
\end{tabular}
\par\end{centering}

\protect\caption{Information on the combinations of $SU(11)$ representations yielding
the correct chirality. The solution $\widetilde{c}+n_{1}-n_{2}$ is
mentioned in \cite{Georgi:1979md}.}
\end{table}

\par\end{center}

\begin{center}
\begin{table}[h]
\begin{centering}
\begin{tabular}{cc}
\toprule 
\begin{tabular}{c}
Maximum size of\tabularnewline
representations considered\tabularnewline
\end{tabular} & 4000\tabularnewline
 & \tabularnewline
\begin{tabular}{c}
Non-trivial combinations\tabularnewline
of representations with\tabularnewline
no chirality ($n_{i}$)\tabularnewline
\end{tabular} & $\begin{array}{c}
-21\repb{12}+6\repb{66}-\rep{220}\,,\\
-64\repb{12}+14\repb{66}-\rep{495}\,,\\
-70\repb{12}+14\repb{66}-\rep{792}\,,\\
28\repb{12}-28\repb{78}+7\repb{572}-\rep{780}-\rep{2145}\,,\\
84\repb{12}-7\repb{78}+\rep{572}-7\repb{780}+\rep{2574}\,.
\end{array}$\tabularnewline
 & \tabularnewline
\begin{tabular}{c}
Interesting solutions \tabularnewline
with the SM chirality\tabularnewline
\end{tabular} & $\begin{array}{rl}
\widetilde{c}= & -24\left(\mathbf{12}\right)+3\left(\mathbf{66}\right)\,,\\
\widetilde{c}-n_{1}= & -3\left(\mathbf{12}\right)-3\left(\mathbf{66}\right)+\mathbf{220}\,.
\end{array}$\tabularnewline
\bottomrule
\end{tabular}
\par\end{centering}

\protect\caption{\label{tab:SU12}Information on the combinations of $SU(12)$ representations
yielding the correct chirality. The solution $\widetilde{c}+4n_{1}-6n_{2}+4n_{3}$
is mentioned in \cite{Albright:2012zt}.}
\end{table}

\par\end{center}


\clearpage

\begin{thebibliography}{10}
\providecommand{\url}[1]{\texttt{#1}}
\providecommand{\urlprefix}{URL }
\providecommand{\eprint}[2][]{\url{#2}}

\bibitem{Georgi:1974sy}
H.~Georgi and S.~Glashow, \emph{{Unity of all elementary-particle forces}},
  \MYhref[journalLinks]{http://dx.doi.org/10.1103/PhysRevLett.32.438}{Phys.
  Rev. Lett.
  }\MYhref[journalLinks]{http://dx.doi.org/10.1103/PhysRevLett.32.438}{\textbf{32}
  (1974) 438--441}.

\bibitem{Georgi:1974my}
H.~Georgi, \emph{{The state of the art --- gauge theories}},
  \MYhref[journalLinks]{http://dx.doi.org/10.1063/1.2947450}{AIP Conf. Proc.
  }\MYhref[journalLinks]{http://dx.doi.org/10.1063/1.2947450}{\textbf{23}
  (1975) 575--582}.

\bibitem{Fritzsch:1974nn}
H.~Fritzsch and P.~Minkowski, \emph{{Unified interactions of leptons and
  hadrons}},
  \MYhref[journalLinks]{http://dx.doi.org/10.1016/0003-4916(75)90211-0}{Annals
  Phys.
  }\MYhref[journalLinks]{http://dx.doi.org/10.1016/0003-4916(75)90211-0}{\textbf{93}
  (1975) 193--266}.

\bibitem{Gursey:1975ki}
F.~G{\"u}rsey, P.~Ramond and P.~Sikivie, \emph{{A universal gauge theory model
  based on $E_6$}},
  \MYhref[journalLinks]{http://dx.doi.org/10.1016/0370-2693(76)90417-2}{Phys.
  Lett. B
  }\MYhref[journalLinks]{http://dx.doi.org/10.1016/0370-2693(76)90417-2}{\textbf{60}
  (1976) 177}.

\bibitem{GellMann:1980vs}
M.~Gell-Mann, P.~Ramond and R.~Slansky, \emph{{Complex spinors and unified
  theories}}, proceedings of the Supergravity Workshop at Stony Brook  (1979)
  315--321, \MYhref[eprintLinks]{http://arxiv.org/abs/1306.4669}{{\ttfamily
  arXiv:1306.4669 [hep-th]}}.

\bibitem{Wilczek:1981iz}
F.~Wilczek and A.~Zee, \emph{{Families from spinors}},
  \MYhref[journalLinks]{http://dx.doi.org/10.1103/PhysRevD.25.553}{Phys. Rev. D
  }\MYhref[journalLinks]{http://dx.doi.org/10.1103/PhysRevD.25.553}{\textbf{25}
  (1982) 553}.

\bibitem{Fujimoto:1981bv}
Y.~Fujimoto, \emph{{$SO(18)$ unification}},
  \MYhref[journalLinks]{http://dx.doi.org/10.1103/PhysRevD.26.3183}{Phys. Rev.
  D
  }\MYhref[journalLinks]{http://dx.doi.org/10.1103/PhysRevD.26.3183}{\textbf{26}
  (1982) 3183}.

\bibitem{Chang:1985jd}
D.~Chang, T.~Hubsch and R.~Mohapatra, \emph{{Grand unification of three light
  generations}},
  \MYhref[journalLinks]{http://dx.doi.org/10.1103/PhysRevLett.55.673}{Phys.
  Rev. Lett.
  }\MYhref[journalLinks]{http://dx.doi.org/10.1103/PhysRevLett.55.673}{\textbf{55}
  (1985) 673}.

\bibitem{Bagger:1984rk}
J.~Bagger and S.~Dimopoulos, \emph{{$O(18)$ revived: splitting the spinor}},
  \MYhref[journalLinks]{http://dx.doi.org/10.1016/0550-3213(84)90192-5}{Nucl.
  Phys. B
  }\MYhref[journalLinks]{http://dx.doi.org/10.1016/0550-3213(84)90192-5}{\textbf{244}
  (1984) 247}.

\bibitem{Hubsch:1985zn}
T.~Hubsch and P.~B. Pal, \emph{{Economical unification of three families in
  $SO(18)$}},
  \MYhref[journalLinks]{http://dx.doi.org/10.1103/PhysRevD.34.1606}{Phys. Rev.
  D
  }\MYhref[journalLinks]{http://dx.doi.org/10.1103/PhysRevD.34.1606}{\textbf{34}
  (1986) 1606}.

\bibitem{Bagger:1985mf}
J.~Bagger, S.~Dimopoulos, E.~Masso and M.~Reno, \emph{{Experimental
  consequences of family unification}},
  \MYhref[journalLinks]{http://dx.doi.org/10.1103/PhysRevLett.54.2199}{Phys.
  Rev. Lett.
  }\MYhref[journalLinks]{http://dx.doi.org/10.1103/PhysRevLett.54.2199}{\textbf{54}
  (1985) 2199}.

\bibitem{Chang:1985uf}
D.~Chang and R.~N. Mohapatra, \emph{{$SO(18)$ unification of fermion
  generations}},
  \MYhref[journalLinks]{http://dx.doi.org/10.1016/0370-2693(85)91191-8}{Phys.
  Lett. B
  }\MYhref[journalLinks]{http://dx.doi.org/10.1016/0370-2693(85)91191-8}{\textbf{158}
  (1985) 323}.

\bibitem{Bagger:1984gz}
J.~Bagger, S.~Dimopoulos, E.~Masso and M.~Reno, \emph{{A realistic theory of
  family unification}},
  \MYhref[journalLinks]{http://dx.doi.org/10.1016/0550-3213(85)90627-3}{Nucl.
  Phys. B
  }\MYhref[journalLinks]{http://dx.doi.org/10.1016/0550-3213(85)90627-3}{\textbf{258}
  (1985) 565}.

\bibitem{Senjanovic:1984rw}
G.~Senjanovic, F.~Wilczek and A.~Zee, \emph{{Reflections on mirror fermions}},
  \MYhref[journalLinks]{http://dx.doi.org/10.1016/0370-2693(84)90269-7}{Phys.
  Lett. B
  }\MYhref[journalLinks]{http://dx.doi.org/10.1016/0370-2693(84)90269-7}{\textbf{141}
  (1984) 389}.

\bibitem{Ida:1980ea}
M.~Ida, Y.~Kayama and T.~Kitazoe, \emph{{Inclusion of generations in
  $SO(14)$}},
  \MYhref[journalLinks]{http://dx.doi.org/10.1143/PTP.64.1745}{Prog. Theor.
  Phys.
  }\MYhref[journalLinks]{http://dx.doi.org/10.1143/PTP.64.1745}{\textbf{64}
  (1980) 1745}.

\bibitem{Masiero:1982xr}
A.~Masiero, M.~Roncadelli and T.~Yanagida, \emph{{A grand unification of
  flavour with testable predictions}},
  \MYhref[journalLinks]{http://dx.doi.org/10.1016/0370-2693(82)90721-3}{Phys.
  Lett. B
  }\MYhref[journalLinks]{http://dx.doi.org/10.1016/0370-2693(82)90721-3}{\textbf{117}
  (1982) 291}.

\bibitem{Sato:1981ga}
H.~Sato, \emph{{Fermion masses from grand unification with $O(14)$}},
  \MYhref[journalLinks]{http://dx.doi.org/10.1016/0370-2693(81)90301-4}{Phys.
  Lett. B
  }\MYhref[journalLinks]{http://dx.doi.org/10.1016/0370-2693(81)90301-4}{\textbf{101}
  (1981) 233}.

\bibitem{Sato:1980jn}
H.~Sato, \emph{{Unification of generations and the Cabibbo angle}},
  \MYhref[journalLinks]{http://dx.doi.org/10.1103/PhysRevLett.45.1997}{Phys.
  Rev. Lett.
  }\MYhref[journalLinks]{http://dx.doi.org/10.1103/PhysRevLett.45.1997}{\textbf{45}
  (1980) 1997}.

\bibitem{Adler:2002yg}
S.~L. Adler, \emph{{Should $E_8$ SUSY Yang-Mills be reconsidered as a family
  unification model?}},
  \MYhref[journalLinks]{http://dx.doi.org/10.1016/S0370-2693(02)01596-4}{Phys.
  Lett. B
  }\MYhref[journalLinks]{http://dx.doi.org/10.1016/S0370-2693(02)01596-4}{\textbf{533}
  (2002) 121--125},
  \MYhref[eprintLinks]{http://arxiv.org/abs/hep-ph/0201009}{{\ttfamily
  arXiv:hep-ph/0201009}}.

\bibitem{Adler:2014pga}
S.~L. Adler, \emph{{$SU(8)$ family unification with boson-fermion balance}},
  \MYhref[journalLinks]{http://dx.doi.org/10.1142/S0217751X14501309}{Int. J.
  Mod. Phys. A
  }\MYhref[journalLinks]{http://dx.doi.org/10.1142/S0217751X14501309}{\textbf{29}
  (2014) 1450130},
  \MYhref[eprintLinks]{http://arxiv.org/abs/1403.2099}{{\ttfamily
  arXiv:1403.2099 [hep-ph]}}.

\bibitem{Dixon:1985jw}
L.~J. Dixon, J.~A. Harvey, C.~Vafa and E.~Witten, \emph{{Strings on
  orbifolds}},
  \MYhref[journalLinks]{http://dx.doi.org/10.1016/0550-3213(85)90593-0}{Nucl.
  Phys. B
  }\MYhref[journalLinks]{http://dx.doi.org/10.1016/0550-3213(85)90593-0}{\textbf{261}
  (1985) 678--686}.

\bibitem{Greene:1986bm}
B.~R. Greene, K.~H. Kirklin, P.~J. Miron and G.~G. Ross, \emph{{A
  three-generation superstring model. (I) Compactification and discrete
  symmetries}},
  \MYhref[journalLinks]{http://dx.doi.org/10.1016/0550-3213(86)90057-X}{Nucl.
  Phys. B
  }\MYhref[journalLinks]{http://dx.doi.org/10.1016/0550-3213(86)90057-X}{\textbf{278}
  (1986) 667}.

\bibitem{Greene:1986jb}
B.~R. Greene, K.~H. Kirklin, P.~J. Miron and G.~G. Ross, \emph{{A
  three-generation superstring model. (II) Symmetry breaking and the
  low-energy theory}},
  \MYhref[journalLinks]{http://dx.doi.org/10.1016/0550-3213(87)90662-6}{Nucl.
  Phys. B
  }\MYhref[journalLinks]{http://dx.doi.org/10.1016/0550-3213(87)90662-6}{\textbf{292}
  (1987) 606}.

\bibitem{Ibanez:1987sn}
L.~E. Ibanez, J.~E. Kim, H.~P. Nilles and F.~Quevedo, \emph{{Orbifold
  compactifications with three families of $SU(3) \times SU(2) \times
  U(1)^n$}},
  \MYhref[journalLinks]{http://dx.doi.org/10.1016/0370-2693(87)90255-3}{Phys.
  Lett. B
  }\MYhref[journalLinks]{http://dx.doi.org/10.1016/0370-2693(87)90255-3}{\textbf{191}
  (1987) 282--286}.

\bibitem{Ibanez:1986tp}
L.~E. Ibanez, H.~P. Nilles and F.~Quevedo, \emph{{Orbifolds and Wilson lines}},
  \MYhref[journalLinks]{http://dx.doi.org/10.1016/0370-2693(87)90066-9}{Phys.
  Lett. B
  }\MYhref[journalLinks]{http://dx.doi.org/10.1016/0370-2693(87)90066-9}{\textbf{187}
  (1987) 25--32}.

\bibitem{Babu:2002ti}
K.~Babu, S.~Barr and B.~Kyae, \emph{{Family unification in five and six
  dimensions}},
  \MYhref[journalLinks]{http://dx.doi.org/10.1103/PhysRevD.65.115008}{Phys.
  Rev. D
  }\MYhref[journalLinks]{http://dx.doi.org/10.1103/PhysRevD.65.115008}{\textbf{65}
  (2002) 115008},
  \MYhref[eprintLinks]{http://arxiv.org/abs/hep-ph/0202178}{{\ttfamily
  arXiv:hep-ph/0202178}}.

\bibitem{Hwang:2002hg}
K.~Hwang and J.~E. Kim, \emph{{Orbifolded $SU(7)$ and unification of
  families}},
  \MYhref[journalLinks]{http://dx.doi.org/10.1016/S0370-2693(02)02150-0}{Phys.
  Lett. B
  }\MYhref[journalLinks]{http://dx.doi.org/10.1016/S0370-2693(02)02150-0}{\textbf{540}
  (2002) 289--294},
  \MYhref[eprintLinks]{http://arxiv.org/abs/hep-ph/0205093}{{\ttfamily
  arXiv:hep-ph/0205093}}.

\bibitem{Choi:2002fn}
K.-S. Choi and J.~E. Kim, \emph{{$Z_2$ orbifold compactification of heterotic
  string and 6D $SO(16)$ and $E_7 \times SU(2)$ flavor unification models}},
  \MYhref[journalLinks]{http://dx.doi.org/10.1016/S0370-2693(02)03104-0}{Phys.
  Lett. B
  }\MYhref[journalLinks]{http://dx.doi.org/10.1016/S0370-2693(02)03104-0}{\textbf{552}
  (2003) 81--86},
  \MYhref[eprintLinks]{http://arxiv.org/abs/hep-th/0206099}{{\ttfamily
  arXiv:hep-th/0206099}}.

\bibitem{Gogoladze:2003ci}
I.~Gogoladze, Y.~Mimura and S.~Nandi, \emph{{Unification of gauge, Higgs and
  matter in extra dimensions}},
  \MYhref[journalLinks]{http://dx.doi.org/10.1016/S0370-2693(03)00564-1}{Phys.
  Lett. B
  }\MYhref[journalLinks]{http://dx.doi.org/10.1016/S0370-2693(03)00564-1}{\textbf{562}
  (2003) 307--315},
  \MYhref[eprintLinks]{http://arxiv.org/abs/hep-ph/0302176}{{\ttfamily
  arXiv:hep-ph/0302176}}.

\bibitem{Gogoladze:2003yw}
I.~Gogoladze, Y.~Mimura and S.~Nandi, \emph{{Unity of elementary particles and
  forces in higher dimensions}},
  \MYhref[journalLinks]{http://dx.doi.org/10.1103/PhysRevLett.91.141801}{Phys.
  Rev. Lett.
  }\MYhref[journalLinks]{http://dx.doi.org/10.1103/PhysRevLett.91.141801}{\textbf{91}
  (2003) 141801},
  \MYhref[eprintLinks]{http://arxiv.org/abs/hep-ph/0304118}{{\ttfamily
  arXiv:hep-ph/0304118}}.

\bibitem{Han:2004qd}
Z.-y. Han and W.~Skiba, \emph{{Family unification on an orbifold}},
  \MYhref[journalLinks]{http://dx.doi.org/10.1103/PhysRevD.70.035013}{Phys.
  Rev. D
  }\MYhref[journalLinks]{http://dx.doi.org/10.1103/PhysRevD.70.035013}{\textbf{70}
  (2004) 035013},
  \MYhref[eprintLinks]{http://arxiv.org/abs/hep-ph/0405199}{{\ttfamily
  arXiv:hep-ph/0405199}}.

\bibitem{Kawamura:2007cm}
Y.~Kawamura, T.~Kinami and K.-y. Oda, \emph{{Orbifold family unification}},
  \MYhref[journalLinks]{http://dx.doi.org/10.1103/PhysRevD.76.035001}{Phys.
  Rev. D
  }\MYhref[journalLinks]{http://dx.doi.org/10.1103/PhysRevD.76.035001}{\textbf{76}
  (2007) 035001},
  \MYhref[eprintLinks]{http://arxiv.org/abs/hep-ph/0703195}{{\ttfamily
  arXiv:hep-ph/0703195}}.

\bibitem{Kawamura:2009gr}
Y.~Kawamura and T.~Miura, \emph{{Orbifold family unification in $SO(2N)$ gauge
  theory}},
  \MYhref[journalLinks]{http://dx.doi.org/10.1103/PhysRevD.81.075011}{Phys.
  Rev. D
  }\MYhref[journalLinks]{http://dx.doi.org/10.1103/PhysRevD.81.075011}{\textbf{81}
  (2010) 075011},
  \MYhref[eprintLinks]{http://arxiv.org/abs/0912.0776}{{\ttfamily
  arXiv:0912.0776 [hep-ph]}}.

\bibitem{Choi:2010gx}
K.-S. Choi and J.~E. Kim, \emph{{Supersymmetric three family chiral $SU(6)$
  grand unification model from F-theory}},
  \MYhref[journalLinks]{http://dx.doi.org/10.1103/PhysRevD.83.065016}{Phys.
  Rev. D
  }\MYhref[journalLinks]{http://dx.doi.org/10.1103/PhysRevD.83.065016}{\textbf{83}
  (2011) 065016},
  \MYhref[eprintLinks]{http://arxiv.org/abs/1012.0847}{{\ttfamily
  arXiv:1012.0847 [hep-ph]}}.

\bibitem{Goto:2013jma}
Y.~Goto, Y.~Kawamura and T.~Miura, \emph{{Orbifold family unification on six
  dimensions}},
  \MYhref[journalLinks]{http://dx.doi.org/10.1103/PhysRevD.88.055016}{Phys.
  Rev. D
  }\MYhref[journalLinks]{http://dx.doi.org/10.1103/PhysRevD.88.055016}{\textbf{88}
  (2013) 5 055016},
  \MYhref[eprintLinks]{http://arxiv.org/abs/1307.2631}{{\ttfamily
  arXiv:1307.2631 [hep-ph]}}.

\bibitem{Frampton:1979cw}
P.~Frampton, \emph{{$SU(N)$ grand unification with several quark--lepton
  generations}},
  \MYhref[journalLinks]{http://dx.doi.org/10.1016/0370-2693(79)90472-6}{Phys.
  Lett. B
  }\MYhref[journalLinks]{http://dx.doi.org/10.1016/0370-2693(79)90472-6}{\textbf{88}
  (1979) 299}.

\bibitem{Georgi:1979md}
H.~Georgi, \emph{{Towards a Grand Unified Theory of flavor}},
  \MYhref[journalLinks]{http://dx.doi.org/10.1016/0550-3213(79)90497-8}{Nucl.
  Phys. B
  }\MYhref[journalLinks]{http://dx.doi.org/10.1016/0550-3213(79)90497-8}{\textbf{156}
  (1979) 126--134}.

\bibitem{Baaklini:1980gd}
N.~Baaklini, \emph{{Chiral grand unification in $SU(N>5)$}},
  \MYhref[journalLinks]{http://dx.doi.org/10.1088/0305-4616/6/8/003}{J. Phys. G
  }\MYhref[journalLinks]{http://dx.doi.org/10.1088/0305-4616/6/8/003}{\textbf{6}
  (1980) 917--931}.

\bibitem{Frampton:1979tj}
P.~Frampton, \emph{{Unification of flavor}},
  \MYhref[journalLinks]{http://dx.doi.org/10.1016/0370-2693(80)90140-9}{Phys.
  Lett. B
  }\MYhref[journalLinks]{http://dx.doi.org/10.1016/0370-2693(80)90140-9}{\textbf{89}
  (1980) 352}.

\bibitem{Kim:1980ci}
J.~E. Kim, \emph{{Model of flavor unity}},
  \MYhref[journalLinks]{http://dx.doi.org/10.1103/PhysRevLett.45.1916}{Phys.
  Rev. Lett.
  }\MYhref[journalLinks]{http://dx.doi.org/10.1103/PhysRevLett.45.1916}{\textbf{45}
  (1980) 1916}.

\bibitem{Kashibayashi:1982nu}
S.~Kashibayashi, K.~Fukuma and T.~Ueda, \emph{{Further constraints on $SU(N)$
  flavor grand unification}},
  \MYhref[journalLinks]{http://dx.doi.org/10.1103/PhysRevD.26.2529}{Phys. Rev.
  D
  }\MYhref[journalLinks]{http://dx.doi.org/10.1103/PhysRevD.26.2529}{\textbf{26}
  (1982) 2529--2531}.

\bibitem{Frampton:1989fu}
P.~H. Frampton and B.-H. Lee, \emph{{$SU(15)$ grand unification}},
  \MYhref[journalLinks]{http://dx.doi.org/10.1103/PhysRevLett.64.619}{Phys.
  Rev. Lett.
  }\MYhref[journalLinks]{http://dx.doi.org/10.1103/PhysRevLett.64.619}{\textbf{64}
  (1990) 619}.

\bibitem{Chkareuli:1992kd}
J.~Chkareuli, \emph{{The SU(8) GUT for chiral families}},
  \MYhref[journalLinks]{http://dx.doi.org/10.1016/0370-2693(93)91346-O}{Phys.
  Lett. B
  }\MYhref[journalLinks]{http://dx.doi.org/10.1016/0370-2693(93)91346-O}{\textbf{300}
  (1993) 361--366}.

\bibitem{Berezhiani:1995dt}
Z.~Berezhiani, \emph{{SUSY $SU(6)$: GIFT for doublet-triplet splitting and
  fermion masses}},
  \MYhref[journalLinks]{http://dx.doi.org/10.1016/0370-2693(95)00705-P}{Phys.
  Lett. B
  }\MYhref[journalLinks]{http://dx.doi.org/10.1016/0370-2693(95)00705-P}{\textbf{355}
  (1995) 481--491},
  \MYhref[eprintLinks]{http://arxiv.org/abs/hep-ph/9503366}{{\ttfamily
  arXiv:hep-ph/9503366}}.

\bibitem{Chkareuli:2000bm}
J.~Chkareuli, C.~Froggatt, I.~Gogoladze and A.~Kobakhidze, \emph{{From
  prototype $SU(5)$ to realistic $SU(7)$ SUSY GUT}},
  \MYhref[journalLinks]{http://dx.doi.org/10.1016/S0550-3213(00)00660-X}{Nucl.
  Phys. B
  }\MYhref[journalLinks]{http://dx.doi.org/10.1016/S0550-3213(00)00660-X}{\textbf{594}
  (2001) 23--45},
  \MYhref[eprintLinks]{http://arxiv.org/abs/hep-ph/0003007}{{\ttfamily
  arXiv:hep-ph/0003007}}.

\bibitem{Shafi:2001iu}
Q.~Shafi and Z.~Tavartkiladze, \emph{{Realistic supersymmetric $SU(6)$}},
  \MYhref[journalLinks]{http://dx.doi.org/10.1016/S0370-2693(01)01255-2}{Phys.
  Lett. B
  }\MYhref[journalLinks]{http://dx.doi.org/10.1016/S0370-2693(01)01255-2}{\textbf{522}
  (2001) 102--106},
  \MYhref[eprintLinks]{http://arxiv.org/abs/hep-ph/0105140}{{\ttfamily
  arXiv:hep-ph/0105140}}.

\bibitem{Barr:2008pn}
S.~Barr, \emph{{Doubly lopsided mass matrices from unitary unification}},
  \MYhref[journalLinks]{http://dx.doi.org/10.1103/PhysRevD.78.075001}{Phys.
  Rev. D
  }\MYhref[journalLinks]{http://dx.doi.org/10.1103/PhysRevD.78.075001}{\textbf{78}
  (2008) 075001},
  \MYhref[eprintLinks]{http://arxiv.org/abs/0804.1356}{{\ttfamily
  arXiv:0804.1356 [hep-ph]}}.

\bibitem{Dent:2009pd}
J.~B. Dent, R.~Feger, T.~W. Kephart and S.~Nandi, \emph{{Natural fermion mass
  hierarchy and mixings in family unification}},
  \MYhref[journalLinks]{http://dx.doi.org/10.1016/j.physletb.2011.02.028}{Phys.
  Lett. B
  }\MYhref[journalLinks]{http://dx.doi.org/10.1016/j.physletb.2011.02.028}{\textbf{697}
  (2011) 367--369},
  \MYhref[eprintLinks]{http://arxiv.org/abs/0908.3915}{{\ttfamily
  arXiv:0908.3915 [hep-ph]}}.

\bibitem{CarcamoHernandez:2010im}
A.~C{\'a}rcamo~Hern{\'a}ndez and R.~Rahman, \emph{{A SUSY $SU(6)$ GUT model
  with pseudo-Goldstone Higgs doublets}}  (2010),
  \MYhref[eprintLinks]{http://arxiv.org/abs/1007.0447}{{\ttfamily
  arXiv:1007.0447 [hep-ph]}}.

\bibitem{Albright:2012zt}
C.~H. Albright, R.~P. Feger and T.~W. Kephart, \emph{{An explicit $SU(12)$
  family and flavor unification model with natural fermion masses and
  mixings}},
  \MYhref[journalLinks]{http://dx.doi.org/10.1103/PhysRevD.86.015012}{Phys.
  Rev. D
  }\MYhref[journalLinks]{http://dx.doi.org/10.1103/PhysRevD.86.015012}{\textbf{86}
  (2012) 015012},
  \MYhref[eprintLinks]{http://arxiv.org/abs/1204.5471}{{\ttfamily
  arXiv:1204.5471 [hep-ph]}}.

\bibitem{Byakti:2013uya}
P.~Byakti, D.~Emmanuel-Costa, A.~Mazumdar and P.~B. Pal, \emph{{Number of
  fermion generations from a novel grand unified model}},
  \MYhref[journalLinks]{http://dx.doi.org/10.1140/epjc/s10052-014-2730-9}{Eur.
  Phys. J. C
  }\MYhref[journalLinks]{http://dx.doi.org/10.1140/epjc/s10052-014-2730-9}{\textbf{74}
  (2014) 2730}, \MYhref[eprintLinks]{http://arxiv.org/abs/1308.4305}{{\ttfamily
  arXiv:1308.4305 [hep-ph]}}.

\bibitem{GellMann:1976pg}
M.~Gell-Mann, P.~Ramond and R.~Slansky, \emph{{Color embeddings, charge
  assignments, and proton stability in unified gauge theories}},
  \MYhref[journalLinks]{http://dx.doi.org/10.1103/RevModPhys.50.721}{Rev. Mod.
  Phys.
  }\MYhref[journalLinks]{http://dx.doi.org/10.1103/RevModPhys.50.721}{\textbf{50}
  (1978) 721}.

\bibitem{King:1981bg}
R.~King, \emph{{Fundamental fermion representations in generalizations of the
  $SU(5)$ Grand Unified Gauge Theory}},
  \MYhref[journalLinks]{http://dx.doi.org/10.1016/0550-3213(81)90368-0}{Nucl.
  Phys. B
  }\MYhref[journalLinks]{http://dx.doi.org/10.1016/0550-3213(81)90368-0}{\textbf{185}
  (1981) 133--156}.

\bibitem{Fonseca:2011sy}
R.~M. Fonseca, \emph{{Calculating the renormalisation group equations of a SUSY
  model with Susyno}},
  \MYhref[journalLinks]{http://dx.doi.org/10.1016/j.cpc.2012.05.017}{Comput.
  Phys. Commun.
  }\MYhref[journalLinks]{http://dx.doi.org/10.1016/j.cpc.2012.05.017}{\textbf{183}
  (2012) 2298--2306},
  \MYhref[eprintLinks]{http://arxiv.org/abs/1106.5016}{{\ttfamily
  arXiv:1106.5016 [hep-ph]}}.

\bibitem{Feger:2012bs}
R.~Feger and T.~W. Kephart, \emph{{LieART -- a Mathematica application for Lie
  algebras and representation theory}}  (2012),
  \MYhref[eprintLinks]{http://arxiv.org/abs/1206.6379}{{\ttfamily
  arXiv:1206.6379 [math-ph]}}.

\bibitem{Dynkin:1957_1}
E.~Dynkin, \emph{{Maximal subgroups of the classical groups}}, Trans. Am. Math.
  Soc. \textbf{6} (1957) 245--378.

\bibitem{Dynkin:1957_2}
E.~Dynkin, \emph{{Semisimple subalgebras of semisimple Lie algebras}}, Trans.
  Am. Math. Soc. \textbf{6} (1957) 111--244.

\bibitem{Losev:2010}
I.~Losev, \emph{{On invariants of a set of elements of a semisimple Lie
  algebra}}, Journal of Lie Theory \textbf{20} (2010) 1 017--030,
  \MYhref[eprintLinks]{http://arxiv.org/abs/math/0512538}{{\ttfamily
  arXiv:math/0512538 [math.RT]}}.

\bibitem{Minchenko:2006}
A.~N.~M. Minchenko, \emph{{The semisimple subalgebras of exceptional Lie
  algebras}},
  \MYhref[journalLinks]{http://dx.doi.org/10.1090/S0077-1554-06-00156-7}{Trans.
  Moscow Math. Soc.
  }\MYhref[journalLinks]{http://dx.doi.org/10.1090/S0077-1554-06-00156-7}{\textbf{67}
  (2006) 225--259}.

\bibitem{Slansky:1981yr}
R.~Slansky, \emph{{Group theory for unified model building}},
  \MYhref[journalLinks]{http://dx.doi.org/10.1016/0370-1573(81)90092-2}{Phys.
  Rept.
  }\MYhref[journalLinks]{http://dx.doi.org/10.1016/0370-1573(81)90092-2}{\textbf{79}
  (1981) 1--128}.

\bibitem{Barr:1981qv}
S.~M. Barr, \emph{{A new symmetry breaking pattern for $SO(10)$ and proton
  decay}},
  \MYhref[journalLinks]{http://dx.doi.org/10.1016/0370-2693(82)90966-2}{Phys.
  Lett. B
  }\MYhref[journalLinks]{http://dx.doi.org/10.1016/0370-2693(82)90966-2}{\textbf{112}
  (1982) 219}.

\bibitem{Bertolini:2010yz}
S.~Bertolini, L.~Di~Luzio and M.~Malinsk{\'y}, \emph{{Minimal flipped $SO(10)
  \otimes U(1)$ supersymmetric Higgs model}},
  \MYhref[journalLinks]{http://dx.doi.org/10.1103/PhysRevD.83.035002}{Phys.
  Rev. D
  }\MYhref[journalLinks]{http://dx.doi.org/10.1103/PhysRevD.83.035002}{\textbf{83}
  (2011) 035002},
  \MYhref[eprintLinks]{http://arxiv.org/abs/1011.1821}{{\ttfamily
  arXiv:1011.1821 [hep-ph]}}.

\bibitem{Dvali:2007hz}
G.~Dvali, \emph{{Black holes and large $N$ species solution to the hierarchy
  problem}},
  \MYhref[journalLinks]{http://dx.doi.org/10.1002/prop.201000009}{Fortsch.
  Phys.
  }\MYhref[journalLinks]{http://dx.doi.org/10.1002/prop.201000009}{\textbf{58}
  (2010) 528--536},
  \MYhref[eprintLinks]{http://arxiv.org/abs/0706.2050}{{\ttfamily
  arXiv:0706.2050 [hep-th]}}.

\bibitem{Calmet:2008df}
X.~Calmet, S.~D. Hsu and D.~Reeb, \emph{{Grand unification and enhanced quantum
  gravitational effects}},
  \MYhref[journalLinks]{http://dx.doi.org/10.1103/PhysRevLett.101.171802}{Phys.
  Rev. Lett.
  }\MYhref[journalLinks]{http://dx.doi.org/10.1103/PhysRevLett.101.171802}{\textbf{101}
  (2008) 171802},
  \MYhref[eprintLinks]{http://arxiv.org/abs/0805.0145}{{\ttfamily
  arXiv:0805.0145 [hep-ph]}}.

\bibitem{Adler:1969gk}
S.~L. Adler, \emph{{Axial-vector vertex in spinor electrodynamics}},
  \MYhref[journalLinks]{http://dx.doi.org/10.1103/PhysRev.177.2426}{Phys. Rev.
  }\MYhref[journalLinks]{http://dx.doi.org/10.1103/PhysRev.177.2426}{\textbf{177}
  (1969) 2426--2438}.

\bibitem{Bell:1969ts}
J.~Bell and R.~Jackiw, \emph{{A PCAC puzzle: $\pi^{0}\rightarrow\gamma\gamma$
  in the $\sigma$-model}},
  \MYhref[journalLinks]{http://dx.doi.org/10.1007/BF02823296}{Nuovo Cim. A
  }\MYhref[journalLinks]{http://dx.doi.org/10.1007/BF02823296}{\textbf{60}
  (1969) 47--61}.

\bibitem{Cahn:1981ub}
R.~N. Cahn, \emph{{Anomalies and the particle content of Grand Unified
  Theories}},
  \MYhref[journalLinks]{http://dx.doi.org/10.1016/0370-2693(81)90126-X}{Phys.
  Lett. B
  }\MYhref[journalLinks]{http://dx.doi.org/10.1016/0370-2693(81)90126-X}{\textbf{104}
  (1981) 282}.

\bibitem{Preskill:1990fr}
J.~Preskill, \emph{{Gauge anomalies in an effective field theory}},
  \MYhref[journalLinks]{http://dx.doi.org/10.1016/0003-4916(91)90046-B}{Annals
  Phys.
  }\MYhref[journalLinks]{http://dx.doi.org/10.1016/0003-4916(91)90046-B}{\textbf{210}
  (1991) 323--379}.

\bibitem{Green:1984sg}
M.~B. Green and J.~H. Schwarz, \emph{{Anomaly cancellation in supersymmetric
  $D=10$ gauge theory and superstring theory}},
  \MYhref[journalLinks]{http://dx.doi.org/10.1016/0370-2693(84)91565-X}{Phys.
  Lett. B
  }\MYhref[journalLinks]{http://dx.doi.org/10.1016/0370-2693(84)91565-X}{\textbf{149}
  (1984) 117--122}.

\bibitem{Singer:1980sw}
M.~Singer, J.~Valle and J.~Schechter, \emph{{Canonical neutral-current
  predictions from the weak-electromagnetic gauge group $SU(3) \times U(1) $}},
  \MYhref[journalLinks]{http://dx.doi.org/10.1103/PhysRevD.22.738}{Phys. Rev. D
  }\MYhref[journalLinks]{http://dx.doi.org/10.1103/PhysRevD.22.738}{\textbf{22}
  (1980) 738}.

\bibitem{Pisano:1991ee}
F.~Pisano and V.~Pleitez, \emph{{$SU(3)\otimes U(1)$ model for electroweak
  interactions}},
  \MYhref[journalLinks]{http://dx.doi.org/10.1103/PhysRevD.46.410}{Phys. Rev. D
  }\MYhref[journalLinks]{http://dx.doi.org/10.1103/PhysRevD.46.410}{\textbf{46}
  (1992) 410--417},
  \MYhref[eprintLinks]{http://arxiv.org/abs/hep-ph/9206242}{{\ttfamily
  arXiv:hep-ph/9206242}}.

\bibitem{Frampton:1992wt}
P.~Frampton, \emph{{Chiral dilepton model and the flavor question}},
  \MYhref[journalLinks]{http://dx.doi.org/10.1103/PhysRevLett.69.2889}{Phys.
  Rev. Lett.
  }\MYhref[journalLinks]{http://dx.doi.org/10.1103/PhysRevLett.69.2889}{\textbf{69}
  (1992) 2889--2891}.

\bibitem{Foot:1992rh}
R.~Foot, O.~F. Hernandez, F.~Pisano and V.~Pleitez, \emph{{Lepton masses in an
  $SU(3)_{L}\otimes U(1)_{N}$ gauge model}},
  \MYhref[journalLinks]{http://dx.doi.org/10.1103/PhysRevD.47.4158}{Phys. Rev.
  D
  }\MYhref[journalLinks]{http://dx.doi.org/10.1103/PhysRevD.47.4158}{\textbf{47}
  (1993) 4158--4161},
  \MYhref[eprintLinks]{http://arxiv.org/abs/hep-ph/9207264}{{\ttfamily
  arXiv:hep-ph/9207264}}.

\bibitem{Boucenna:2014dia}
S.~M. Boucenna, R.~M. Fonseca, F.~Gonz{\'a}lez-Canales and J.~W.~F. Valle,
  \emph{{Small neutrino masses and gauge coupling unification}},
  \MYhref[journalLinks]{http://dx.doi.org/10.1103/PhysRevD.91.031702}{Phys.
  Rev. D
  }\MYhref[journalLinks]{http://dx.doi.org/10.1103/PhysRevD.91.031702}{\textbf{91}
  (2015) 3 031702(R)},
  \MYhref[eprintLinks]{http://arxiv.org/abs/1411.0566}{{\ttfamily
  arXiv:1411.0566 [hep-ph]}}.

\end{thebibliography}
\end{document}